\documentclass[12pt]{article}
\usepackage{indentfirst} 
\usepackage{amsfonts,amssymb,amsmath}
\usepackage{bm}          
\usepackage{cite}
\usepackage{url}
\usepackage{enumitem}
\usepackage[usenames,dvipsnames]{xcolor}
\usepackage{todonotes}
\usepackage{multicol}
\usepackage{MnSymbol}
\usepackage{hyperref}
\usepackage{tensor}
\usepackage{arcs}
\usepackage[titletoc,title]{appendix}

\usetikzlibrary{arrows}

\AtBeginDocument{}%

\newcommand{\corurl}{red}
\newcommand{\corcite}{ForestGreen}
\newcommand{\corlink}{blue}
\newcommand{\iprod}{\raisebox{8pt}{\scalebox{1}[-2.5]{$\neg\,$}}}

\newcommand{\dd}{\mathrm{d\!l}}

\newcommand{\id}{\mathrm{Id}}

\hypersetup{linktocpage,colorlinks,urlcolor=\corurl,citecolor=\corcite,linkcolor=\corlink,
pdftitle={Euclidean self-dual gravity: Ashtekar variables without gauge fixing},
pdfauthor={J.F. Barbero, M. Basquens, E.J.S. Villase\~nor}}

\numberwithin{equation}{section}  

\def\QED{{\boldmath$\rule{0.5em}{0.5em}$}}                                
\def\markatright#1{\leavevmode\unskip\nobreak\quad\hspace*{\fill}{#1}}    
\def\qed{\markatright{\QED}}                                              

\usepackage{titling}
\usepackage{authblk}
\title{Euclidean self-dual gravity: Ashtekar variables without gauge fixing}
\author[1,3]{J. Fernando Barbero G.}
\author[2,4]{Marc Basquens\,}
\author[2,3]{Eduardo J.S. Villase\~nor\,}
\affil[1]{Instituto de Estructura de la Materia, IEM-CSIC. Serrano 123, 28006 Madrid, Spain}
\affil[2]{Departamento de Matem\'aticas, Universidad Carlos III de Madrid, Avda.\  de la Universidad 30, 28911 Legan\'es, Spain}
\affil[3]{Grupo de Teor\'{\i}as de Campos y F\'{\i}sica Estad\'{\i}stica. Instituto Gregorio Mill\'an (UC3M). Unidad Asociada al Instituto de Estructura de la Materia, CSIC}
\affil[4]{Department of Energy Technology, Royal Institute of Technology (KTH), 10044 Stockholm, Sweden}
\date{}                     
\begin{document}
	
\maketitle
\date{March 31, 2023}


\begin{abstract}
The GNH method is used in this paper to study the Hamiltonian formulation of the Euclidean self-dual action. This action can be used to arrive at the complex Ashtekar formulation of General Relativity or a real connection formulation for Euclidean General Relativity. The main result of the paper is a derivation of the Ashtekar formulation for Euclidean gravity without using any gauge fixing. It is interesting to compare this derivation with the one corresponding to the Holst action. In particular it is worth noting that no ``tertiary'' constraints appear in the case considered in the present paper.
\end{abstract}

\tableofcontents


%
%
\section{Introduction}{\label{sec_intro}}

The self-dual action \cite{Samuel:1987td,Jacobson:1987yw} played a very important role in the establishment of the Ashtekar formulation for General Relativity (GR) as a new starting point for the quantization of gravity. The self-dual formulation has also drawn interest in the context of the quantization proposal for GR put forward by Smolin in \cite{smolin} and discussed recently in \cite{M1,M2,T1,T2,nos}. At variance with the Hamiltonian treatment of the Hilbert-Palatini action---which leads to a version of standard geometrodynamics endowed with an $SO(3)$ internal symmetry---the self-dual version of the Hilbert-Palatini action leads to the Ashtekar formulation for either complex or real-Euclidean GR. Although at the present moment the standard way to arrive at the real Ashtekar formulation makes use of the Holst action \cite{Holst:1995pc}, the self-dual action is still widely considered as a means to address a number of problems, in particular whenever the complex formulation is useful (see, for instance, \cite{Wieland:2017zkf,Wieland:2023qzp}). For this reason it is interesting to revisit the Hamiltonian formulation derived from it and consider several issues that have not completely cleared out in the past, in particular regarding the stability of all the constraints when the Dirac method is used \cite{BarberoG:1994fcp}. A very powerful way to address these issues is to rely on the geometric perspective provided by the GNH approach to Hamiltonian dynamics \cite{Gotay1,Gotay2}. In this setting the main consistency condition is translated into a tangency requirement whose verification does not need the computation of Poisson brackets and, thus, is significantly easier than its counterpart in the Dirac ``algorithm''.

It is interesting to compare the GNH analysis of the Holst action (discussed, for instance, in \cite{Rezende:2009sv} from the standard Dirac point of view and also in \cite{BarberoG:2021ekv} from the perspective of the GNH method). As we show here, on one hand, the analysis of the self-dual action is shorter because no new secondary constraints appear after the first batch of them show up in the analysis, whereas, in the Holst case, secondary constraints appear at two different stages. On the other hand, the explicit verification of the tangency of the Hamiltonian vector fields to the phase space submanifold determined by the secondary constraints in the case of the self-dual action is rather intricate and requires careful consideration.

One interesting consequence of the analysis that we present in the paper is the possibility of deriving the Ashtekar formulation for Euclidean GR \textit{without any gauge fixing}. The key insight to arrive at this result comes from the form of the pullback of the canonical symplectic form to the primary constraint submanifold. In fact it is possible to write this pullback in a way that leads to the introduction of a function in the phase space that takes the sympletic form to a simple ``canonical'' form (i.e. the analogous of $\Sigma_i \mathrm{d}q_i\wedge \mathrm{d}p^i$). It is very interesting at this point to rewrite the remaining elements of the Hamiltonian formulation (constraints and Hamiltonian vector fields) with the help of this new object. It is also worth to compare the result with the one derived by the usual time-gauge fixing. As we will show the results are compatible in an appealing way and the role of one of the $SO(3)$  factors of the $SO(4)$ symmetry is noteworthy.

The structure of the paper is the following. After this introduction we will quickly discuss some basic facts about the self-dual (actually the anti-delf-dual) action for Euclidean GR in section \ref{sec_Action}. After that we will discuss in Section \ref{sec_Hamiltonian} the Hamiltonian formulation for this action and perform the first steps of the GNH analysis. Before completing the crucial tangency tests to verify the consistency of the dynamics obtained by using the GNH method we will discuss some features of the secondary constraints (Section
\ref{sec_constraints}) and solve the equations that determine the components of the Hamiltonian vector fields on the primary constraint submanifold (Section \ref{sec_vector_fields}). With all this information we will give a detailed account in Section \ref{sec_tangency} of the tangency analysis, although we will leave some details for Appendix \ref{appendix_details}. Although the main ideas involved are simple, the computation itself is quite intricate. In Section \ref{sec_summary} we show how the Ashtekar formulation appears quite naturally by following our approach, in fact, this is probably the cleanest way to arrive at it from an action principle. Furthermore, as we will explain there is no need in principle to use any gauge fixing (at variance with the situation in the case of the Holst action, see, for instance \cite{BarberoG:2020tit,BarberoG:2021ekv}). We end the paper with some conclusions in Section \ref{sec_conclusions}. Appendix \ref{appendix_Gauss} gives some computations related to the Gauss law in the Ashtekar formulation.

Some comments about our notation. As a general rule we will employ boldface characters to denote four dimensional geometric objects and non-boldface letters for the 3-dimensional ones. The totally antisymmetric Levi-Civita symbol in three dimensions will be denoted as $\epsilon_{ijk}$. The ``internal'' indices $i,j,k,...$ will be raised and lowered with the Euclidean metric $\mathrm{Diag}(+++)$ (so, in practice, their position upstairs or downstairs is irrelevant).  We will also employ boldface characters to denote canonical momenta. The exterior differentials in $\mathcal{M}$, $\Sigma$ and the phase space $T^*Q$  will be respectively denoted as $\bm{\mathrm{d}}$, $\mathrm{d}$ and $\mathrm{d\!l}$. The interior product of a vector field $X$ and a differential form $\beta$ will be denoted either as $\imath_X\beta$ or $X\iprod\beta$. Throughout the paper $\pounds_X$ denotes the Lie derivative along a vector field $X$. Finally, the scalar field $\phi$ satisfying $\sigma=\phi\,\mathsf{vol}$ for a top-form $\sigma$ and a volume form $\mathsf{vol}$ will be often written as $\left(\frac{\sigma}{\mathsf{vol}}\right)$.

%
%

\section{Action and equations of motion}\label{sec_Action}

Let $\Sigma$ be a closed (i.e. compact without boundary), orientable, 3-dimensional manifold (this implies that $\Sigma$ is paralellizable) and $\mathcal{M}=\mathbb{R}\times\Sigma$. The basic fields that we will use to write the action are:
\begin{align*}
{\bm{\mathrm{e}}}^i\in\Omega^1(\mathcal{M})\,,&&i=1,2,3\\
{\bm{\omega}}^i\in\Omega^1(\mathcal{M})\,,&&i=1,2,3\\
{\bm{\alpha}}\in\Omega^1(\mathcal{M})\,.
\end{align*}
The fields ${\bm{\alpha}}$ and ${\bm{\mathrm{e}}}^i$ are chosen in such a way that ${\bm{\alpha}}\otimes{\bm{\alpha}}+{\bm{\mathrm{e}}}_i\otimes {\bm{\mathrm{e}}}^i$ is a Euclidean metric of signature $(+\,+\,+\,\,+)$. As a consequence $({\bm{\alpha}},{\bm{\mathrm{e}}}^i)$ defines a non-degenerate tetrad.

Important geometric objects defined with the help of the fields introduced above are the covariant exterior differential ${\bm{\mathrm{D}}}$, which acts on the ${\bm{\mathrm{e}}}_i$ according to
\[
{\bm{\mathrm{D}}}{\bm{\mathrm{e}}}_i:={\bm{\mathrm{d}}}{\bm{\mathrm{e}}}_i+\varepsilon_{ijk}{\bm{\omega}}^j\wedge {\bm{\mathrm{e}}}^k\,,
\]
and the curvature 2-form
\[
{\bm{\mathrm{F}}}^i:={\bm{\mathrm{d}}}{\bm{\omega}}^i+\frac{1}{2}\varepsilon^{ijk}{\bm{\omega}}_j\wedge{\bm{\omega}}_k\,.
\]
The Euclidean self-dual action for GR can be written in the form \cite{BarberoG:1994fcp}
\begin{equation}\label{self_dual_action}
  S({\bm{\mathrm{e}}},{\bm{\omega}},{\bm{\alpha}}):=\int_{\mathcal{M}} \left(\frac{1}{2}\varepsilon_{ijk}{\bm{\mathrm{e}}}^i\wedge {\bm{\mathrm{e}}}^j\wedge {\bm{\mathrm{F}}}^k-{\bm{\alpha}}\wedge {\bm{\mathrm{e}}}_i\wedge {\bm{\mathrm{F}}}^i \right)\,.
\end{equation}
Notice that the first term is the Husain-Kucha\v{r} action \cite{Husain:1990vz}. The indices $i,j,k=1,2,3$ can be considered as ``$SO(3)$ indices'' because \eqref{self_dual_action} is invariant under the infinitesimal gauge transformations
\begin{align}\label{SO3transformations_1}
  \begin{split}
  \delta_1 {\bm{\omega}}^i& = {\bm{\mathrm{D}}}{\bm{\Lambda}}^i\,,\\
  \delta_1 {\bm{\alpha}}& =0\,, \\
  \delta_1 {\bm{\mathrm{e}}}^i& =\varepsilon^i_{\phantom{i}jk}{\bm{\mathrm{e}}}^j{\bm{\Lambda}}^k\,,
  \end{split}
\end{align}
with ${\bm{\Lambda}}^k\in C^\infty(\mathcal{M})$. The action is also invariant under the infinitesimal transformations
\begin{align}\label{SO3transformations_2}
  \begin{split}
  \delta_2 {\bm{\omega}}^i& = 0\,,\\
  \delta_2 {\bm{\alpha}}& ={\bm{\Upsilon}}_i {\bm{\mathrm{e}}}^i\,, \\
  \delta_2 {\bm{\mathrm{e}}}^i& =-{\bm{\Upsilon}}^i{\bm{\alpha}}+\varepsilon^i_{\phantom{i}jk}{\bm{\mathrm{e}}}^j{\bm{\Upsilon}}^k\,,
  \end{split}
\end{align}
where ${\bm{\Upsilon}}^k\in C^\infty(\mathcal{M})$. It is important to point out that $\delta_1$ and $\delta_2$ are \textit{independent}. Also, it is worth noting that these transformation do not commute, in fact they satisfy
\[
[\delta_1({\bm{\Lambda}}),\delta_2({\bm{\Upsilon}})]=\delta_2({\bm{\Lambda}}\times {\bm{\Upsilon}})\,.
\]
The internal symmetries of the action can be written in other forms, for instance
\begin{equation}\label{dual_symmetry}
\left\{\begin{array}{l}
\delta^-({\bm{\Lambda}}){\bm{\omega}}_i={\bm{D}}{\bm{\Lambda}}_i\\
\delta^-({\bm{\Lambda}}){\bm{\alpha}}=-\frac{1}{2}{\bm{\Lambda}}_i{\bm{\mathrm{e}}}^i\\
\delta^-({\bm{\Lambda}}){\bm{\mathrm{e}}}_i=\frac{1}{2}{\bm{\Lambda}}_i{\bm{\alpha}}+\frac{1}{2}\varepsilon_{ijk}{\bm{\mathrm{e}}}^j{\bm{\Lambda}}^k
\end{array}\right.\,,\quad\left\{
\begin{array}{l}
\delta^+({\bm{\Upsilon}}){\bm{\omega}}_i=0\\
\delta^+({\bm{\Upsilon}}){\bm{\alpha}}=\frac{1}{2}{\bm{\Upsilon}}_i{\bm{\mathrm{e}}}^i\\
\delta^+({\bm{\Upsilon}}){\bm{\mathrm{e}}}_i=-\frac{1}{2}{\bm{\Upsilon}}_i{\bm{\alpha}}+\frac{1}{2}\varepsilon_{ijk}{\bm{\mathrm{e}}}^j{\bm{\Upsilon}}^k
\end{array}
\right.\,.
\end{equation}
At variance with $\delta_1$ and $\delta_2$ these transformations \textit{do commute}, i.e.
\[
[\delta^-({\bm{\Lambda}}),\delta^+({\bm{\Upsilon}})]=0\,.
\]
Taking also into account that
\[
[\delta^-({\bm{\Lambda}}),\delta^-({\bm{M}})]=\delta^-({\bm{\Lambda}}\times{\bm{M}})\,,\quad [\delta^+({\bm{\Lambda}}),\delta^+({\bm{M}})]=\delta^+({\bm{\Lambda}}\times{\bm{M}})\,,
\]
we see that \eqref{dual_symmetry} provides an explicit realization of $SO(4)=SO(3)\otimes SO(3)$ as  the symmetry group of the self-dual action \eqref{self_dual_action} \cite{BarberoG:1994fcp}. Notice that
\[
\delta_1({\bm{\Lambda}})=\delta^-({\bm{\Lambda}})+\delta^+({\bm{\Lambda}})\,,\quad \delta_2({\bm{\Upsilon}})=\delta^+(2{\bm{\Upsilon}})\,.
\]
The field equations coming from the action \eqref{self_dual_action} are
\begin{align}\label{field_equations}
   & {\bm{\mathrm{D}}}({\bm{\alpha}\wedge \bm{\mathrm{e}}_k})+\epsilon_{ijk}\bm{\mathrm{e}}^i\wedge {\bm{\mathrm{D}}} \bm{\mathrm{e}}^j=0\,, \\
   & \epsilon_{ijk}\bm{\mathrm{e}}^j\wedge \bm{\mathrm{F}}^k+\bm{\alpha}\wedge \bm{\mathrm{F}}_i=0\,,\\
   & \bm{\mathrm{e}}^i\wedge \bm{\mathrm{F}}_i=0\,.
\end{align}
They are equivalent to the Euclidean Einstein equations in vacuum.

\section{The Hamiltonian formulation in the GNH approach}\label{sec_Hamiltonian}

In order to get the Lagrangian and Hamiltonian formulations given by the self-dual action \eqref{self_dual_action} we take advantage of the foliation naturally associated with $\mathcal{M}=\mathbb{R}\times\Sigma$. The spatial sheets of this foliation are $\Sigma_t:=\{t\}\times\Sigma$ ($t\in\mathbb{R}$). The foliation also defines a canonical evolution vector field $\partial_t$ given by the tangent vectors to the curves $c_p:\mathbb{R}\rightarrow \mathbb{R}\times\Sigma:\tau\mapsto (\tau,p)$ with $p\in\Sigma$. For each $t\in\mathbb{R}$  we define the embedding $\jmath_t:\Sigma\rightarrow\mathcal{M}:p\mapsto(t,p)$ and denote its pullback as $\jmath_t^\ast$. Notice that $\Sigma_t=\jmath_t(\Sigma)$.

By remembering that
\[
\int_{\mathbb{R}\times\Sigma}\mathcal{L}=\int_{\mathbb{R}\times\Sigma}\mathrm{d}t\wedge\imath_{\partial_t}\mathcal{L}=\int_{\mathbb{R}}\mathrm{d}t\int_{\Sigma_t}\imath_{\partial_t}\mathcal{L}
=\int_{\mathbb{R}}\mathrm{d}t\int_{\Sigma}\jmath_t^*\imath_{\partial_t}\mathcal{L}\,.
\]
we can compute the Lagrangian $L:=\int_\Sigma\jmath_t^*\imath_{\partial_t}\mathcal{L}$ defined on $\Sigma$ from the 4-form $\mathcal{L}$ appearing in the action. The Lagrangian thus obtained is defined on the configuration space
\[
Q=\mathcal{C}^\infty(\Sigma)^3 \times \Omega^1(\Sigma)^3 \times\mathcal{C}^\infty(\Sigma)^3 \times \Omega^1(\Sigma)^3\times\mathcal{C}^\infty(\Sigma) \times \Omega^1(\Sigma)\,,
\]
with elements of the form $(e_{\mathrm{t}}^i,e^i,\omega_{\mathrm{t}}^i,\omega^i,\alpha_{\mathrm{t}},\alpha)$, by interpreting the objects
\begin{align*}
&  e_{\mathrm{t}}^i(t)\,:=\jmath_t^\ast\imath_{\partial_t}\bm{\mathrm{e}}^i\,, && e^i(t)\,:=\jmath_t^\ast \bm{\mathrm{e}}^i\,,\\
&  \omega_{\mathrm{t}}^i(t):=\jmath_t^\ast\imath_{\partial_t}\bm{\omega}^i\,, && \omega^i(t):=\jmath_t^\ast \bm{\omega}^i\,,\\
&  \alpha_{\mathrm{t}}(t):=\jmath_t^\ast\imath_{\partial_t}\bm{\alpha}\,, && \alpha(t)\,\,:=\jmath_t^\ast \bm{\alpha}\,.
\end{align*}
as defining curves in the configuration space $Q$ and considering also their velocities
\begin{align*}
  & v_{e_{\mathrm{t}}}^i(t):=\jmath_t^\ast \pounds_{\partial_t} (\imath_{\partial_t}{\bm{\mathrm{e}}}^i)=\frac{\mathrm{d}e_{\mathrm{t}}^i}{\mathrm{d}\tau}(t)\,, && v_{e}^i(t):=\jmath_t^\ast \pounds_{\partial_t} \bm{\mathrm{e}}^i=\frac{\mathrm{d}}{\mathrm{d}\tau}(\jmath_\tau^\ast \bm{\mathrm{e}}^i)\Big{|}_{\tau=t}=\frac{\mathrm{d}e^i}{\mathrm{d}\tau}(t)\,,\\
  & v_{\omega_{\mathrm{t}}}^i(t):=\jmath_t^\ast \pounds_{\partial_t} (\imath_{\partial_t}{\bm{\mathrm{\omega}}}^i)=\frac{\mathrm{d}\omega_{\mathrm{t}}^i}{\mathrm{d}\tau}(t)\,, && v_{\omega}^i(t):=\jmath_t^\ast \pounds_{\partial_t} \bm{\mathrm{\omega}}^i=\frac{\mathrm{d}}{\mathrm{d}\tau}(\jmath_\tau^\ast \bm{\mathrm{\omega}}^i)\Big{|}_{\tau=t}=\frac{\mathrm{d}\omega^i}{\mathrm{d}\tau}(t)\,,\\
  & v_{\alpha_{\mathrm{t}}}(t):=\jmath_t^\ast \pounds_{\partial_t} (\imath_{\partial_t}{\bm{\alpha}})=\frac{\mathrm{d}\alpha_{\mathrm{t}}}{\mathrm{d}\tau}(t)\,, && v_{\alpha}(t):=\jmath_t^\ast \pounds_{\partial_t} \bm{\alpha}=\frac{\mathrm{d}}{\mathrm{d}\tau}(\jmath_\tau^\ast {\bm\alpha})\Big{|}_{\tau=t}=\frac{\mathrm{d}\alpha}{\mathrm{d}\tau}(t)\,,
\end{align*}
defined in terms of the Lie derivative $\pounds_{\partial_t}$ along the vector field $\partial_t$. Here we are leaving aside functional issues---necessary for the complete definition of the configuration space---that play an important role in the rigorous Hamiltonian formulation of this model. In any case, the results that we will obtain here are in par (and equivalent to) with those derived by using Dirac's method. The Lagrangian is
\begin{align*}
    L(\mathrm{v}) = &\int_\Sigma \Big( \left(\frac{1}{2} \epsilon_{ijk} e^i \wedge e^j +e_k\wedge\alpha \right) \wedge v_\omega^k +  \omega_{\mathrm{t}}^i D \left( \frac{1}{2}\epsilon_{ijk} e^j \wedge e^k +  e_i\wedge\alpha \right)\\
    & \hspace{6.7cm}- \alpha_{\mathrm{t}} e_i \wedge F^i + e_{\mathrm{t}}^i \left( \epsilon_{ijk} e^j \wedge F^k + \alpha \wedge F^i \right) \Big)\,,
\end{align*}
where we denote the velocity in the fiber of $TQ$ corresponding to $(e_{\mathrm{t}}^i,e^i,\omega_{\mathrm{t}}^i,\omega^i,\alpha_{\mathrm{t}},\alpha)$ as $\mathrm{v}:=(v_{e_\mathrm{t}}^i, v_e^i,v_{\omega_\mathrm{t}}^i, v_\omega^i,v_{\alpha_\mathrm{t}},v_\alpha)$ (as a consequence $L:TQ\rightarrow \mathbb{R}$ depends both on the components of the velocity and the fields defining the configuration manifold $Q$). We also define $F^i:=d\omega^i+\frac{1}{2}\epsilon^{ijk}\omega_j\wedge\omega_k$ and $De^i:=de^i+\epsilon^{ijk}\omega_j\wedge e_k$ (with the usual generalization to forms of other degrees). An important consequence of the non-degeneracy of the tetrad $(\bm{\alpha},{\bm{\mathrm{e}}}^i)$ is the non-degeneracy of the triad $e^i$ on any $\Sigma_t$.

The fiber derivative that defines the canonical momenta is given by
\begin{equation}\label{FL}
    FL(\mathrm{v})(\mathrm{w})=\int_\Sigma \left(\frac{1}{2}\epsilon_{ijk} e^i \wedge e^j +  e_k \wedge \alpha \right) \wedge w_\omega^k\,,
\end{equation}
hence, the primary constraint submanifold $\mathsf{M}_0$ in the phase space is defined by the conditions

\vspace*{-3mm}

\begin{alignat*}{3}
    &\mathbf{p}_{e_{\mathrm{t}}} &&= 0 \,, \\
    &\mathbf{p}_e &&= 0 \,, \\
    &\mathbf{p}_{\omega_{\mathrm{t}}} &&= 0 \,, \\
    &\mathbf{p}_\omega (\mathrm{w}) &&= \int_\Sigma \left( \frac{1}{2}\epsilon_{ijk} e^i \wedge e^j  +  e_k \wedge \alpha \right) \wedge w_\omega^k \,, \\
    &\mathbf{p}_{\alpha_{\mathrm{t}}} &&= 0 \,,\\
    &\mathbf{p}_\alpha &&= 0 \,.
\end{alignat*}

The Hamiltonian is defined on the primary constraint submanifold by $H=E\circ FL^{-1}$ where $E:=\langle FL(v),v\rangle-L$ is the energy (a real function in $TQ$). As in this case the energy $E$ only depends on the configuration variables the functional form of the Hamiltonian coincides with that of $E$, hence
\begin{equation}
\label{asd_hamiltonian}
     H(\mathbf{p}) = \int_\Sigma \Big(\alpha_{\mathrm{t}} e_i \wedge F^i- \omega_{\mathrm{t}}^i D \left( \frac{1}{2}\epsilon_{ijk} e^j \wedge e^k +e_i\wedge \alpha\right)  - e_{\mathrm{t}}^i \left( \epsilon_{ijk} e^j \wedge F^k + \alpha \wedge F_i \right) \Big)\,.
\end{equation}
Here we denote the momenta in the fiber of $T^*Q$ over the point $(e_{\mathrm{t}}^i,e^i,\omega_{\mathrm{t}}^i,\omega^i,\alpha_{\mathrm{t}},\alpha)\in Q$ as $\mathbf{p}:=(\mathbf{p}_{e_{\mathrm{t}}},\mathbf{p}_e,\mathbf{p}_{\omega_{\mathrm{t}}},\mathbf{p}_\omega,\mathbf{p}_{\alpha_{\mathrm{t}}},\mathbf{p}_\alpha)$. The Hamiltonian depends on $\mathbf{p}$ only through its base point.

\medskip

A vector field in phase space $\mathbb{Y}\in\mathfrak{X}(T^*Q)$ has the following component structure
\[
\mathbb{Y}=(Y_{e_{\mathrm{t}}}^i, Y_e^i, Y_{\omega_{\mathrm{t}}}^i, Y_\omega^i, Y_{\alpha_{\mathrm{t}}}, Y_\alpha, {\bm{\mathrm{Y}}}_{\!\!\bm{\mathrm{p}}_{e_{\mathrm{t}}}}, {\bm{\mathrm{Y}}}_{\!\!\bm{\mathrm{p}}_{e}}, {\bm{\mathrm{Y}}}_{\!\!\bm{\mathrm{p}}_{\omega_{\mathrm{t}}}}, {\bm{\mathrm{Y}}}_{\!\!\bm{\mathrm{p}}_{\omega}}, {\bm{\mathrm{Y}}}_{\!\!\bm{\mathrm{p}}_{\alpha_{\mathrm{t}}}}, {\bm{\mathrm{Y}}}_{\!\!\bm{\mathrm{p}}_{\alpha}})\,,
\]
where we use boldface characters for those components of $\mathbb{Y}$ that are dual objects. Notice that $Y_e^i\,,Y_\omega^i\,,Y_\alpha\in \Omega^1(\Sigma)$ and $Y_{e_{\mathrm{t}}}^i\,,Y_{\omega_{\mathrm{t}}}^i\,,Y_{\alpha_{\mathrm{t}}}\in \mathcal{C}^\infty(\Sigma)$.

We can then write the action of the exterior differential in phase space $\dd H$ acting on a vector field $\mathbb{Y}$ as
\begin{align}
  \dd H(\mathbb{Y})=\int_\Sigma\Big(&-Y_{e_{\mathrm{t}}}^i(\epsilon_{ijk}e^j\wedge F^k+\alpha\wedge F_i)\nonumber  \\
   & +Y_e^i\wedge \big(\alpha_{\mathrm{t}}F_i+\epsilon_{ijk}(D\omega_{\mathrm{t}}^j)\wedge e^k-D\omega_{\mathrm{t}}^i\wedge \alpha+\epsilon_{ijk}e_{\mathrm{t}}^jF^k\big)\nonumber \\
   & +Y_{\omega_{\mathrm{t}}}^i D(\alpha\wedge e_ i-\frac{1}{2}\epsilon_{ijk}e^j\wedge e^k) \label{dH}\\
   & +Y_\omega^i\wedge\big(D(\alpha_{\mathrm{t}}e_i+\epsilon_{ijk}e_je_{\mathrm{t}}^k-\alpha e_{\mathrm{t}}^i)-\omega_{\mathrm{t}}^je_j\wedge e_ i-\epsilon_{ijk}\omega_{\mathrm{t}}^j\alpha\wedge e^k\big) \nonumber\\
   & +Y_{\alpha_{\mathrm{t}}} e_i\wedge F^i \nonumber\\
   & +Y_\alpha\wedge\big(D\omega_{\mathrm{t}}^i\wedge e_i-e_{\mathrm{t}}^iF_i\big)\Big)\,.\nonumber
\end{align}

\medskip

The action of the canonical symplectic form $\Omega$ on $\mathbb{Y}, \mathbb{Z}\in \mathfrak{X}(T^*Q)$ can be written in terms of the components of these vector fields as
\begin{align*}
  \Omega(\mathbb{Z},\mathbb{Y}) & ={\bm{\mathrm{Y}}}_{\!\!\bm{\mathrm{p}}_{e_{\mathrm{t}}}}(Z_{e_{\mathrm{t}}})  -  {\bm{\mathrm{Z}}}_{\bm{\mathrm{p}}_{e_{\mathrm{t}}}}(Y_{e_{\mathrm{t}}})
                                  +{\bm{\mathrm{Y}}}_{\!\!\bm{\mathrm{p}}_{e}}(Z_{e})  -  {\bm{\mathrm{Z}}}_{\bm{\mathrm{p}}_{e}}(Y_{e})
                                  +{\bm{\mathrm{Y}}}_{\!\!\bm{\mathrm{p}}_{\omega_{\mathrm{t}}}}(Z_{\omega_{\mathrm{t}}})  -  {\bm{\mathrm{Z}}}_{\bm{\mathrm{p}}_{\omega_{\mathrm{t}}}}(Y_{\omega_{\mathrm{t}}})\\
                                & +{\bm{\mathrm{Y}}}_{\!\!\bm{\mathrm{p}}_{\omega}}(Z_{\omega})  -  {\bm{\mathrm{Z}}}_{\bm{\mathrm{p}}_{\omega}}(Y_{\omega})
                                  +{\bm{\mathrm{Y}}}_{\!\!\bm{\mathrm{p}}_{\alpha_{\mathrm{t}}}}(Z_{\alpha_{\mathrm{t}}})  -  {\bm{\mathrm{Z}}}_{\bm{\mathrm{p}}_{\alpha_{\mathrm{t}}}}(Y_{\alpha_{\mathrm{t}}})
                                  +{\bm{\mathrm{Y}}}_{\!\!\bm{\mathrm{p}}_{\alpha}}(Z_{\alpha})  -  {\bm{\mathrm{Z}}}_{\bm{\mathrm{p}}_{\alpha}}(Y_{\alpha})\,.
\end{align*}
The vector fields $\mathbb{Y}\in \mathfrak{X}(T^*Q)$ tangent to the primary constraint submanifold $\mathsf{M}_0$ have the form
\[
\mathbb{Y}=(Y_{e_{\mathrm{t}}}, Y_e, Y_{\omega_{\mathrm{t}}}, Y_\omega, Y_{\alpha_{\mathrm{t}}}, Y_\alpha,{\bm{0}},{\bm{0}},{\bm{0}},{\bm{\mathrm{Y}}},{\bm{0}},{\bm{0}})
\]
with
\[
{\bm{\mathrm{Y}}}(\cdot)=\int_\Sigma (\epsilon_{ijk}Y_e^i\wedge e^j-\alpha\wedge Y_e^k-Y_\alpha\wedge e_k)\wedge \cdot
\]

One of the central ideas in the GNH approach is to work directly on the primary constraint submanifold $\mathsf{M}_0$. In the present case this is especially easy as $\mathsf{M}_0$ is \textit{precisely} the configuration space $Q$. This means that vector fields on this manifold have the form
\[
\mathbb{Y}_0=(Y_{e_{\mathrm{t}}}^i, Y_e^i, Y_{\omega_{\mathrm{t}}}^i, Y_\omega^i, Y_{\alpha_{\mathrm{t}}}, Y_\alpha)\,.
\]
The pullback of $\Omega$ to $\mathsf{M}_0$---another of the basic elements of the GNH approach---can be written as
\begin{align}
&\omega(\mathbb{Z}_0,\mathbb{Y}_0)=\int_\Sigma\Big(Y_e^i\wedge(\epsilon_{ijk}e^j\wedge Z_\omega^k+\alpha\wedge Z_{\omega i})\label{pullback_omega}\\
&\hspace*{5.8cm}+Y_\omega^i\wedge(Z_\alpha\wedge e_i-\epsilon_{ijk}Z_e^j\wedge e^k-Z_e^i\wedge \alpha)+Y_\alpha\wedge Z_\omega^i\wedge e_i\Big)\,.\nonumber
\end{align}
The basic equation of the GNH approach $\imath_{\mathbb{Z}_0}\omega=\dd H$ can now be easily solved by equating the terms proportional to the different components of the vector field $\mathbb{Y}$ in \eqref{dH} and \eqref{pullback_omega}. By doing this one finds the secondary constraints
\begin{align}
&\epsilon_{ijk}e^j\wedge F^k+\alpha\wedge F_i=0\,,\label{L1}\\
&D(\frac{1}{2}\epsilon_{ijk}e^j\wedge e^k+e_i\wedge\alpha)=0\,,\label{L2}\\
&e_ i\wedge F^i=0\,,\label{L3}
\end{align}
and the following equations for the components of the Hamiltonian vector field $\mathbb{Z}_0$
\begin{align}
  & (\epsilon_{ijk}e^j+\delta_{ik}\alpha)\wedge(Z_\omega^k-D\omega_{\mathrm{t}}^k)=(\delta_{ik}\alpha_{\mathrm{t}}+\epsilon_{ijk}e_{\mathrm{t}}^j)F^k\,,\label{E1}\\
  & (\epsilon_{ijk}e^j-\delta_{ik}\alpha)\wedge(Z_e^k-De_{\mathrm{t}}^k-\epsilon^k_{\phantom{k}\ell m}e^\ell\omega_{\mathrm{t}}^m)+e_i\wedge(Z_\alpha-\mathrm{d}\alpha_{\mathrm{t}})\label{E2}\\
  & \hspace*{8cm}=e_{\mathrm{t}}^i\mathrm{d}\alpha+(\epsilon_{ijk}e_{\mathrm{t}}^j-\alpha_{\mathrm{t}}\delta_{ik})De^k\,\nonumber\\
  & e_i\wedge(Z_\omega^i-D \omega_{\mathrm{t}}^i)=e_{\mathrm{t}}^iF_ i\,.\label{E3}
\end{align}
There are no conditions on $Z^i_{e_{\mathrm{t}}}$, $Z^i_{\omega_{\mathrm{t}}}$ and $Z_{\alpha_{\mathrm{t}}}$ so they are arbitrary and, hence, the dynamics of $e_{\mathrm{t}}^i$, $\omega_{\mathrm{t}}^i$ and $\alpha_{\mathrm{t}}$ is also arbitrary.

The next step in the GNH method is to check if the vector fields whose components satisfy \eqref{E1}-\eqref{E3} are tangent to the submanifold of $\mathsf{M}_0$ defined by the secondary constraints \eqref{L1}-\eqref{L3}. Before performing this analysis---which turns out to be rather intricate---it is necessary to analyse in detail the secondary constraints and also to solve for $Z_e^i$, $Z_\omega^i$ and $Z_\alpha$ in \eqref{E1}-\eqref{E3}. We devote the next two sections to this issue.

%
%
\section{Some details on the secondary constraints}\label{sec_constraints}

In order to study the tangency of the Hamiltonian vector fields given by \eqref{E1}-\eqref{E3} it is necessary to disentangle part of the content of the constraints \eqref{L1} and \eqref{L3}. We do it here. First we point out that the curvature $F_i$ can be written as
\[
F_i=\frac{1}{2}\mathbb{F}_{ij}\epsilon^{jk\ell}e_k\wedge e_\ell
\]
with
\[
\mathbb{F}_{ij}:=\left(\frac{F_i\wedge e_j}{\mathsf{vol}_e}\right)\,.
\]
Here $\mathsf{vol}_e:=\frac{1}{3!}\epsilon_{ijk}e^i\wedge e^j\wedge e^k$ is a volume form on $\Sigma$ because the $e^i$ are linearly independent everywhere.

As the $e^i$ are linearly independent we can write $\alpha=\alpha_i e^i$ and put the secondary constraint \eqref{L1} in the form
\[
\epsilon_{ijk}e^j\wedge F^k+\alpha\wedge F_i=(\mathbb{F}_{ij}\alpha^j-\epsilon_{ijk}\mathbb{F}^{jk})\mathsf{vol}_e=0\,,
\]
which is equivalent to
\begin{equation}\label{L1_new}
\mathbb{F}_{ij}\alpha^j-\epsilon_{ijk}\mathbb{F}^{jk}=0\,.
\end{equation}
It is also straightforward to rewrite \eqref{L3} in terms of $\mathbb{F}_{ij}$, indeed
\[
e^i\wedge F_i=\frac{1}{2}e_i\wedge \mathbb{F}^i_{\,\,j}\epsilon^{jk\ell}e_k\wedge e_\ell=\frac{1}{2}\epsilon_{ik\ell}\epsilon^{jkl}\mathbb{F}^i_{\,\,j}\mathsf{vol}_e=\mathbb{F}^i_{\,\,i}\mathsf{vol}_e\,,
\]
and, hence, the secondary constraint \eqref{L3} is equivalent to
\begin{equation}\label{L3_new}
\mathbb{F}^i_{\,\,i}=0\,.
\end{equation}
If we expand now $\mathbb{F}_{ij}$ in terms of a symmetric-traceless part $S_{ij}$, a trace $S$ and an antisymmetric part $\epsilon_{ijk}A_k$
\[
\mathbb{F}_{ij}=S_{ij}+\frac{1}{3}\delta_{ij}S+\epsilon_{ijk} A^k\,,
\]
the constraint \eqref{L3_new} implies $S=0$ and \eqref{L1_new} becomes
\[
(2\delta_{ij}+\epsilon_{ijk}\alpha^k)A^j=S_{ij}\alpha^j\,,
\]
which can be solved for $A_i$ in terms of the symmetric-traceless object $S_{ij}$
\[
A_i=\frac{1}{2(4+\alpha^2)}(4\delta_{ij}+\alpha_i\alpha_j-2\epsilon_{ijk}\alpha^k)S^j_{\,\,\ell}\alpha^\ell\,,
\]
where $\alpha^2:=\alpha_i\alpha^i$. We then conclude that, when the constraints hold
\begin{equation}\label{Fij}
\mathbb{F}_{ij}=S_{ij}+\frac{1}{2(4+\alpha^2)}\epsilon_{ijk}(4\delta^k_{\,\,\ell}+\alpha^k\alpha_{\ell}-2\epsilon^k_{\,\,\ell m}\alpha^m)S^\ell_{\,\,p}\alpha^p\,,
\end{equation}
with $S_{ij}$ symmetric and traceless. Notice that this is not the full solution to the constraints \eqref{L1}-\eqref{L3} but, rather, a restriction on the form of the curvature $F_i$. In order to fully solve the constraits one should find the connection $\omega_i$ from \eqref{Fij}.

%
%
\section{The Hamiltonian vector fields}\label{sec_vector_fields}

In this section we will solve equations \eqref{E1}-\eqref{E3} for $Z_e^i$, $Z_\omega^i$ and $Z_\alpha$. For this purpose it helps to define
\begin{align*}
  X_e^k & :=Z_e^k-De_{\mathrm{t}}^k-\epsilon^k_{\phantom{k}\ell m}e^\ell\omega_{\mathrm{t}}^m\,, \\
  X_\omega^k & :=Z_\omega^k-D\omega_{\mathrm{t}}^k\,,\\
  X_\alpha & :=Z_\alpha-\mathrm{d}\alpha_{\mathrm{t}}\,.
\end{align*}
so that \eqref{E1}-\eqref{E3} become
\begin{align}
  & (\epsilon_{ijk}e^j+\delta_{ik}\alpha)\wedge X_\omega^k=(\delta_{ik}\alpha_{\mathrm{t}}+\epsilon_{ijk}e_{\mathrm{t}}^j)F^k\,,\label{EX1}\\
  & (\epsilon_{ijk}e^j-\delta_{ik}\alpha)\wedge X_e^k+e_i\wedge X_\alpha=e_{\mathrm{t}i}\mathrm{d}\alpha+(\epsilon_{ijk}e_{\mathrm{t}}^j-\alpha_{\mathrm{t}}\delta_{ik})De^k\,\label{EX2}\\
  & e_i\wedge X_\omega^i=e_{\mathrm{t}}^iF_ i\,.\label{EX3}
\end{align}
We solve now these equations taking into account that the secondary constraints \eqref{L1}-\eqref{L3} hold.
\subsection{Solving equation \ref{EX1}}\label{subsec_EX1}
By expanding the 1-forms $X_\omega^k=W^k_{\phantom{k}q}\,e^q$ and $\alpha=\alpha_i e^i$ we can write \eqref{EX1} as the following linear, inhomogeneous equation for $W_{ij}$
\begin{equation}\label{EW1}
  W_{ij}-\delta_{ij}W^k_{\phantom{k}k}-\epsilon_i^{\phantom{i}pq}\alpha_pW_{jq}+\alpha_{\mathrm{t}}\mathbb{F}_{ji}+\epsilon_j^{\phantom{i}pq}e_{\mathrm{t}p}\mathbb{F}_{qi}=0\,.
\end{equation}
We expand now $W_{ij}$ in irreducible components as
\begin{equation}\label{expansion_W}
W_{ij}=w_{ij}+\frac{1}{3}\delta_{ij}w+\epsilon_{ijk}w^k\,,
\end{equation}
with $w_{ij}$ symmetric and traceless. From \eqref{expansion_W} we see that
\begin{equation}\label{ws}
w^i=\frac{1}{2}\epsilon^{ijk}W_{jk}\,,\quad w=W^i_{\phantom{i}i}\,.
\end{equation}
Multiplying \eqref{EW1} by $\epsilon^{kij}$ and $\alpha^i$ respectively we get
\begin{align*}
&\epsilon^{ijk}W_{jk}-\alpha_j W^{ji}+\alpha^i W^j_{\phantom{j}j}-\alpha_{\mathrm{t}}\epsilon^{ijk}\mathbb{F}_{jk}-\mathbb{F}^i_{\phantom{i}j}e_{\mathrm{t}}^j=0\\
&\alpha_jW^{ji}-\alpha^i W^j_{\phantom{j}j}+\alpha_{\mathrm{t}}\mathbb{F}_{ij}\alpha^j+\epsilon_{ijk}e_{\mathrm{t}}^j\mathbb{F}^{k\ell}\alpha_\ell=0\,.
\end{align*}
Adding both expressions, taking into account \eqref{ws} and using the constraints in the form \eqref{L1_new}-\eqref{L3_new} we find
\[
w^i=\frac{1}{2}e_{\mathrm{t}j}\mathbb{F}^{ji}\,.
\]

In order to find $w$ we take the trace of \eqref{EW1}, which gives
\[
-2W^i_{\phantom{i}i}+\epsilon^{ijk}\alpha_ iW_{jk}+\epsilon^{ijk}e_{\mathrm{t}i}\mathbb{F}_{jk}=0\,,
\]
after making use of the constraint $\mathbb{F}^i_{\phantom{i}i}=0$. From the previous expression, the value of $w^i$ computed above and the constraint \eqref{L1_new} we get
\[
w=\epsilon_{ijk}e_{\mathrm{t}}^i\mathbb{F}^{jk}\,.
\]
A crucial simplification of equation \eqref{EW1} can be achieved by using $W^{i}_{\phantom{n}i}=w$, noting that $W_{jq}=W_{qj}-2\epsilon_{qjk}w^k$ and introducing this expression in the third term of \eqref{EW1}. By doing this we find that \eqref{EW1} can be rewritten as
\begin{equation}\label{EW1_bis}
  (\delta_{iq}+\epsilon_{iqp}\alpha^p)W^q_{\phantom{q}j}+\alpha_{\mathrm{t}}\mathbb{F}_{ji}-\alpha_je_{\mathrm{t}k}\mathbb{F}^k_{\phantom{k}i}+\epsilon_{jpq}e_{\mathrm{t}}^p\mathbb{F}^q_{\phantom{q}i}=0\,,
\end{equation}
which can be easily solved by inverting the $2\times2$ matrix $\delta_{iq}+\epsilon_{iqp}\alpha^p$ as the index $j$ in $W^q_{\phantom{q}j}$ is a mere spectator. The result is
\begin{equation}\label{Wij}
  W_{ij}=\frac{1}{1+\alpha^2}\big(\delta_i^{\phantom{n}k}+\alpha_ i\alpha^k-\epsilon_i^{\phantom{n}k\ell}\alpha_\ell\big)M_{kj}\,,
\end{equation}
with
\begin{equation}\label{Mij}
M_{kj}:=\alpha_je_{\mathrm{t}q}\mathbb{F}^q_{\phantom{i}k}-\epsilon_{jpq}e_{\mathrm{t}}^p\mathbb{F}^q_{\phantom{i}k}-\alpha_{\mathrm{t}}\mathbb{F}_{jk}\,.
\end{equation}

\subsection{Solving equation \ref{EX2}}\label{subsec_EX2}

By following exactly the same steps as in the previous subsection it is possible to solve equation \eqref{EX2}. In this case we expand $X_e^k=E^k_{\phantom{k}q}\,e^q$ and $X_\alpha=X_ie^i$. We also take into account that the constraint \eqref{L2} is equivalent to
\[
\mathrm{d}\alpha\wedge e_i=\alpha\wedge De_i-\epsilon_{ijk}e^j\wedge De^k\,.
\]
This means that on the constraint submanifold defined by the secondary constraints we have
\begin{equation}
\mathrm{d}\alpha=\frac{1}{2}\epsilon_{ijk}\left(\frac{\mathrm{d}\alpha\wedge e_i}{\mathsf{vol}_e}\right)e^j\wedge e^k=\left(\frac{1}{2}\mathbb{B}^i_{\phantom{i}\ell}\alpha^\ell\epsilon_{ijk}+\mathbb{B}_{jk}\right)e^j\wedge e^k\,,
\end{equation}
where $\mathbb{B}_{ij}:=\left(\frac{De_i\wedge e_j}{\mathsf{vol}_e}\right)$.

Taking all this into account, on the constraint hypersurface equation \eqref{EX2} can be written as
\begin{equation}\label{EW2}
E_{ij}-\delta_{ij}E^k_{\phantom{k}k}+\epsilon_i^{\phantom{i}pq}\alpha_p E_{jq}-\epsilon_{ijk}X^k-\alpha_{\mathrm{t}}\mathbb{B}_{ji}+\epsilon_{jk\ell}e_{\mathrm{t}}^k\mathbb{B}^\ell_{\phantom{i}i}+\mathbb{B}_{ik}\alpha^ke_{\mathrm{t}j}+\epsilon_i^{\phantom{i}pq}\mathbb{B}_{pq}e_{\mathrm{t}j}=0\,,  \end{equation}
where the $X_i$ can be taken to be completely arbitrary.

The resolution of this equation follows the same steps as that of \eqref{EW1} so we just quote the result here:
\begin{equation}\label{Eij}
  E_{ij}=\frac{1}{1+\alpha^2}\big(\delta_i^{\phantom{n}k}+\alpha_ i\alpha^k+\epsilon_i^{\phantom{n}k\ell}\alpha_\ell\big)N_{kj}\,,
\end{equation}
with
\begin{align}
  N_{ij}&:=&\hspace*{-5mm}\delta_{ij}\Big(&\epsilon_{k\ell m}e_{\mathrm{t}}^k\mathbb{B}^{\ell m}+\epsilon_{k\ell m}\alpha^ke_{\mathrm{t}}^\ell\mathbb{B}^{mn}\alpha_n+\frac{1}{2}(e_{\mathrm{t}}\cdot\alpha)(\alpha\mathbb{B}\alpha)+\frac{1}{2}(e_{\mathrm{t}}\cdot\alpha)\epsilon_{k\ell m}\alpha^k\mathbb{B}^{\ell m}\nonumber\\
  & & &\hspace*{0cm}-\frac{1}{2}(e_{\mathrm{t}}\cdot\alpha)\mathbb{B}+(\alpha\mathbb{B}e_{\mathrm{t}})+\alpha\cdot X-\frac{1}{2}\alpha_{\mathrm{t}}(\alpha\mathbb{B}\alpha)-\frac{1}{2}\alpha_{\mathrm{t}}\epsilon_{k\ell m}\alpha^k\mathbb{B}^{\ell m}-\frac{1}{2}\alpha_{\mathrm{t}}\mathbb{B}\Big)\nonumber\\
  & & &\hspace*{-10mm}+\alpha_j\Big(\mathbb{B} e_{\mathrm{t}i}-2\epsilon_{ipq}e_{\mathrm{t}}^p\mathbb{B}^q_{\phantom{q}r}\alpha^r-e_{\mathrm{t}i}(\alpha \mathbb{B}\alpha)-e_{\mathrm{t}i}\epsilon_{k\ell m}\alpha^k\mathbb{B}^{\ell m}-2\mathbb{B}_{ik}e_{\mathrm{t}}^k+e_{\mathrm{t}k}\mathbb{B}^k_{\phantom{i}i}-2X_i\nonumber\\
  & & &\hspace*{5.7cm}-\epsilon_{i\ell m}\alpha^\ell X^m+\alpha_{\mathrm{t}}\mathbb{B}_{i\ell}\alpha^\ell+\alpha_{\mathrm{t}}\epsilon_{i\ell m}\mathbb{B}^{\ell m}\Big)\nonumber\\
  & & &\hspace*{-10mm}-\epsilon_{jpq}e_{\mathrm{t}}^p\mathbb{B}^q_{\phantom{i}i}-\mathbb{B}_{i\ell}\alpha^\ell e_{\mathrm{t}j}-\epsilon_{ipq}\mathbb{B}^{pq}e_{\mathrm{t}j}+\epsilon_{ijk}X^k+\alpha_{\mathrm{t}}\mathbb{B}_{ji}\,,\label{Nij}
\end{align}
where the following shorthand notation has been used:
\[
e_{\mathrm{t}}\cdot\alpha:=e_{\mathrm{t}i}\alpha^i\,,\,\,\,\alpha\mathbb{B}\alpha:=\alpha_i\mathbb{B}^{ij}\alpha_j\,,\,\,\,\mathbb{B}:=\mathbb{B}^{i}_{\phantom{n}i}   \,,\,\,\,\alpha\mathbb{B}e_{\mathrm{t}}:=\alpha_i\mathbb{B}^{ij}e_{\mathrm{t}j}\,,\,\,\,\alpha\cdot X:=\alpha_i X^i\,.
\]

\subsection{Solving equation \ref{EX3}}\label{subsec_EX3}

Equation \eqref{E3} can be easily written in the form $\epsilon^{ijk}W_{jk}=e_{\mathrm{t}j}\mathbb{F}^{ji}$. It is obvious from the discussion presented in subsection \ref{subsec_EX1} (and straightforward to check) that $W_{ij}$, as given in \eqref{Wij}, satisfies \eqref{E3}.

%
%
\section{Tangency analysis}\label{sec_tangency}

The consistency of the Hamiltonian dynamics requires that the Hamiltonian vector fields obtained by solving \eqref{EX1}-\eqref{EX3} must be tangent to the submanifold of $\mathsf{M}_0$ defined by the secondary constraints \eqref{L1}-\eqref{L3}. These tangency conditions can be easily obtained by computing the derivatives of the functions defining the constraints along the field (i.e. $\imath_{\mathbb{X}}\mathrm{d\!l}$). By doing this one gets
\begin{align}
  & (\epsilon_{ijk}Z_e^j+\delta_{ik}Z_\alpha)\wedge F^k+(\epsilon_{ijk}e^j+\delta_{ik}\alpha)\wedge DZ_\omega^k=0\,,\label{t1}\\
  & D(\epsilon_{ijk}e^j\wedge Z_e^k-Z_\alpha\wedge e_i-\alpha\wedge Z_{ei})+Z_\omega^k\wedge(e^i\wedge e_k-\epsilon^i_{\phantom{i}km}\alpha\wedge e^m)=0\,,\label{t2}\\
  & Z_e^i\wedge F_i+e_i\wedge DZ_\omega^i=0\,.\label{t3}
\end{align}
Before checking if the Hamiltonian vector fields given by \eqref{Wij} and \eqref{Eij} satisfy the conditions \eqref{t1}-\eqref{t3} it is convenient to simplify them, in particular by removing the covariant differential $D$ of the components of the vector field $\mathbb{Z}$. As we show next \eqref{t1}-\eqref{t3} can be written in the more convenient form
\begin{align}
  & \epsilon_{ijk}X_e^j\wedge F^k+X_\alpha\wedge F_i+D(\epsilon_{ijk}e^j+\delta_{ik}\alpha)\wedge X_\omega^k=0\,,\label{tan1}\\
  & (\delta_{ij} e_k-\epsilon_{ijk}\alpha)\wedge e^j\wedge X_\omega^k+(\alpha_{\mathrm{t}}\delta_{ik}+\epsilon_{ijk}e_{\mathrm{t}j})\alpha\wedge F^k+e_{\mathrm{t}}^je_i\wedge F_j=0\,,\label{tan2}\\
  & X_e^i\wedge F_i+X_\omega^i\wedge De_i=0\,.\phantom{\big[}\label{tan3}
\end{align}
In the case of \eqref{t1} one first takes the covariant exterior differential $D$ of \eqref{E1} to get
\begin{align*}
 (\epsilon_{ijk}e_j+\delta_{ik}\alpha)\wedge DZ_\omega^k=&(\epsilon_{ijk}De^j+\delta_{ik}\mathrm{d}\alpha)\wedge(Z_\omega^k-D\omega_{\mathrm{t}}^k)\\
 &+D\omega_{\mathrm{t}m}\epsilon^{mk\ell}(\epsilon_{ijk}e^j+\delta_{ik}\alpha)\wedge F_\ell-(\delta_{ik}\mathrm{d}\alpha_{\mathrm{t}}+\epsilon_{ijk}De_{\mathrm{t}}^j)\wedge F^k\,,
\end{align*}
then introduces this expression into \eqref{t1} and simplifies the result by using the constraints \eqref{L1} and \eqref{L3} to get \eqref{tan1}.

In order to rewrite the tangency condition \eqref{t2} in a simpler way one first writes \eqref{E2} in the form
\[
\epsilon_{ijk}e^j\wedge Z_e^k-Z_\alpha\wedge e_i-\alpha\wedge Z_{ei}=D(\epsilon_{ijk}e_{\mathrm{t}}^je^k+\alpha e_{\mathrm{t}}^i-\alpha_{\mathrm{t}}e^i)+\omega_{\mathrm{t}}^k e_k\wedge e_i-\epsilon^i_{\phantom{i}jk}\alpha_\ell \omega_{\mathrm{t}}^k e^\ell\wedge e^j\,,
\]
and then takes the covariant differential of this expression which, by using the constraints (\ref{L2},\ref{L3}), gives
\[
D(\epsilon_{ijk}e^j\wedge Z_e^k-Z_\alpha\wedge e_i-\alpha\wedge Z_{ei})=\alpha_{\mathrm{t}}\epsilon_{ijk}e^j\wedge F^k-e_{\mathrm{t}}^j(\epsilon_{ijk}\alpha+\delta_{jk}e_i)\wedge F^k+(e_k\delta_{ij}-\epsilon_{ijk}\alpha)\wedge e^j\wedge D\omega_{\mathrm{t}}^k\,.
\]
Introducing this now into \eqref{t2} gives \eqref{tan2}.

Finally, in order to arrive at \eqref{tan3} one first takes the exterior covariant differential of \eqref{E3} to get
\[
e_i\wedge DZ_{\omega}^i=De_i\wedge Z_\omega^i-De_i\wedge D\omega_{\mathrm{t}}^i+\epsilon_{ijk}e^i\wedge F^j\omega_{\mathrm{t}}^k-De_{\mathrm{t}}^i\wedge F_i\,,
\]
and then introduces this expression into \eqref{t3}.

In the remaining of this section we will check that the tangency conditions \eqref{tan1}-\eqref{tan3} hold on the submanifold of $\mathsf{M}_0$ defined by the secondary constraints \eqref{L1}-\eqref{L3}. We will start from the easiest to the hardest ones. As we will see, the computations are quite involved. Although, by necessity, we will have to skip many details we will provide enough information to enable the motivated readers to complete them.

One source of difficulties is the possibility of having many different ways to write a particular expression by making use of the constraints \eqref{L1}-\eqref{L3}. One possible way to avoid this problem is to use \eqref{Fij} and write everything in terms of $S_{ij}$. Although this is possible in principle, in practice the computations are very long. A better strategy---that ultimately works---is to use the fact that the constraints imply $\mathbb{F}_{ij}=S_{ij}+\epsilon_{ijk}A^k$, with $S_{ij}$ symmetric and traceless, and use $S_{ij}\alpha^j=(2\delta_{ij}+\epsilon_{ijk}\alpha^k)A^j$ [equivalent to \eqref{L1_new}], whenever possible, to write everything in terms of $A^k$.

\subsection{Checking condition \ref{tan2}}\label{subsec_tangencia2}

The only tangency condition which is easy to check is \eqref{tan2}. In order to see that it holds it suffices to left-wedge-multiply \eqref{EX1}---the equation that must be solved to obtain $X_\omega^i$---by $\alpha$ to get
\[
\alpha\wedge(\epsilon_{ijk}e^j+\delta_{ik}\alpha)\wedge X_\omega^k=(\delta_{ik}\alpha_{\mathrm{t}}+\epsilon_{ijk}e_{\mathrm{t}}^j)\alpha\wedge F^k\,.
\]
Plugging this into \eqref{tan2} leads to
\[
e_k\wedge e_i\wedge X_\omega^k+e_{\mathrm{t}}^je_i\wedge F_j=0\,.
\]
which can be immediately seen to hold as a consequence of \eqref{EX3}.

\subsection{Checking condition \ref{tan3}}\label{subsec_tangencia3}

In order to check the tangency requirement expressed by \eqref{tan3} we will first rewrite it as an equivalent condition in terms of the $M_{ij}$ and $N_{ij}$ introduced above [see (\ref{Mij}) and (\ref{Nij})]:
\begin{equation*}
\mathbb{B}^{ij}M_{ij}+(\alpha^i \mathbb{B}_{ij})(\alpha_k M^{kj})+\epsilon^{ijk}\alpha_iM_{j\ell}\mathbb{B}_k^{\phantom{k}\ell}+\mathbb{F}^{ij}N_{ij}+(\alpha^i \mathbb{F}_{ij})(\alpha_k N^{kj})-\epsilon^{ijk}\alpha_iN_{j\ell}\mathbb{F}_k^{\phantom{k}\ell}=0\,.
\end{equation*}
In principle one just has to substitute (\ref{Mij}) and (\ref{Nij}) in the previous expression and show that the result is zero. A possible way to do this is to use tensor manipulating packages such as xAct \cite{xAct}. However, in our opinion it is instructive to do the computation by hand as some important simplifications are quite non-trivial. To this end it is helpful to separately consider the terms depending on $X_i$, $\alpha_{\mathrm{t}}$ and $e_{\mathrm{t}}^i$. In the first two instances ($X_i$ and $\alpha_{\mathrm{t}}$) the computations are quite direct and the cancelations of the different terms obvious. The only hints worth mentioning here are
\begin{itemize}
\item Use $\mathbb{F}_{ij}=S_{ij}+\epsilon_{ijk}A^k$, taking into account, whenever necessary, that $S_i^{\phantom{i}i}=0$.
\item Use the constraints in the form \eqref{L1_new} to write $\mathbb{F}^{ij}\alpha_j=2A^i$. Also use $S_{ij}\alpha^j=2A_i+\epsilon_{ijk}A^j\alpha^k$ whenever the combination $S_{ij}\alpha^j$ appears.
\item Replace $\mathbb{B}_{ij}-\mathbb{B}_{ji}$ by $\epsilon_{ijk}\epsilon^{k\ell m}\mathbb{B}_{\ell m}$ whenever possible.
\end{itemize}
By doing this all the terms involving $X_i$, and those proportional to  $\alpha_{\mathrm{t}}$ cancel.

\medskip

The computation of the terms proportional to $e_{\mathrm{t}}^i$ is significantly harder, so we will give more details about it. In this case the terms coming from the direct substitution of (\ref{Mij}) and (\ref{Nij}) into the tangency condition do not cancel automatically in an obvious way. In fact, the result is

\begin{align*}
  & -\epsilon_{ijk}e_{\mathrm{t}}^iS^j_{\phantom{i}\ell}\mathbb{B}^{\ell k}-\epsilon_{ijk}e_{\mathrm{t}}^i\mathbb{B}^{j\ell}S_{\ell}^{\phantom{i}k}
  -\epsilon_{ijk}\mathbb{B}^{ij}S^k_{\phantom{k}\ell}e_{\mathrm{t}}^\ell\hspace*{8cm}
\end{align*}

\vspace*{-.7cm}

\begin{align*}
  & -4\epsilon_{ijk}e_{\mathrm{t}}^i(\mathbb{B}^{j\ell}\alpha_\ell)A^k-3(e_{\mathrm{t}}\cdot A)\epsilon_{ijk}\alpha^i\mathbb{B}^{jk}+3\epsilon_{ijk}\alpha^iA^j(e_{\mathrm{t}\ell}\mathbb{B}^{\ell k})+5\epsilon_{ijk}\alpha^i(\mathbb{B}^{j\ell}e_{\mathrm{t}\ell})A^k\\
  & -2\mathbb{B}\epsilon_{ijk}\alpha^ie_{\mathrm{t}}^jA^k\!\!+\!2\epsilon_{ijk}\alpha^ie_{\mathrm{t}}^j(\mathbb{B}^{k\ell}A_\ell)\!+\!(e_{\mathrm{t}}\cdot\alpha)\epsilon_{ijk}A^i\mathbb{B}^{jk}
  \!\!+\!2(A\cdot\alpha)\epsilon_{ijk}e_{\mathrm{t}}^i\mathbb{B}^{jk}\!\!-\!2\epsilon_{ijk}(\alpha_\ell\mathbb{B}^{\ell i})e_{\mathrm{t}}^jA^k
\end{align*}

\vspace*{-.7cm}

\begin{align*}
  &+2(\alpha\mathbb{B}\alpha)\epsilon_{ijk}e_{\mathrm{t}}^iA^j\alpha^k+2\alpha^2\epsilon_{ijk}A^ie_{\mathrm{t}}^j(\mathbb{B}^{k\ell}\alpha_\ell)-2(e_{\mathrm{t}}\cdot\alpha)\epsilon_{ijk}(\mathbb{B}^{i\ell}\alpha_\ell)A^j\alpha^k\hspace*{3.2cm}\\
  &-2(A\cdot\alpha)\epsilon_{ijk}\alpha^ie_{\mathrm{t}}^j(\mathbb{B}^{k\ell}\alpha_\ell)\hspace*{3.2cm}
\end{align*}

Several features of the previous expression stand out, in particular, all the terms involve $\epsilon_{ijk}$, some of them depend on $S_{ij}$, and the remaining ones are either linear or cubic in $\alpha_i$.

In order to show that the $S$-dependent terms cancel out it suffices to realize that $0=\epsilon_{[ijk}S^j_{\phantom{i}\ell]}$ implies $\epsilon_{ijk}S^j_{\phantom{i}\ell}-\epsilon_{jk\ell}S^j_{\phantom{i}i}-\epsilon_{\ell ij}S^j_{\phantom{i}k}=0$, where we have used $S^i_{\phantom{i}i}=0$.

The strategy to show that the other terms cancel is similar to this but more involved so we leave some of the details for Appendix \ref{appendix_details}.

\subsection{Checking condition \ref{tan1}}\label{subsec_tangencia1}

As in the previous subsection we will write the tangency condition \eqref{tan1} as an equivalent expression in terms of the $M_{ij}$ and $N_{ij}$. To avoid having ``dangling indices'' we will multiply it by an arbitrary object $C_i$ (which can be removed at the end). By doing this we get
\begin{align}
& \phantom{-\,\,}\epsilon^{ijk}C_i\mathbb{B}_j^{\phantom{j}\ell}M_{k\ell}+\epsilon^{ijk}\alpha_iC_j\mathbb{B}_k^{\phantom{k}\ell}(\alpha^m M_{m\ell})-(\alpha_i\mathbb{B}^{ij})(C^kM_{kj})+(C\cdot\alpha)\mathbb{\mathbb{B}}^{ij}M_{ij}\nonumber\\
& -\epsilon^{ijk}C_i\mathbb{F}_j^{\phantom{j}\ell}N_{k\ell}\,-\,\epsilon^{ijk}\alpha_iC_j\mathbb{F}_k^{\phantom{k}\ell}(\alpha^m N_{m\ell})\,-\,(\alpha_i\mathbb{F}^{ij})(C^kN_{kj})\,+(C\cdot\alpha)\mathbb{\mathbb{F}}^{ij}N_{ij}\nonumber\\
& +(C^iM_{ij})(\mathbb{B}^{jk}\alpha_k)+(C\cdot\alpha)(\alpha^iM_{ij})(\mathbb{B}^{jk}\alpha_k)-\epsilon^{ijk}\alpha_iC_jM_{k\ell}(\mathbb{B}^{\ell m}\alpha_m)\nonumber\\
& +(C^iM_{ij})\epsilon^{jk\ell}\mathbb{B}_{k\ell}+(C\cdot\alpha)(\alpha^iM_{ij})\epsilon^{jk\ell}\mathbb{B}_{k\ell}-\epsilon^{ijk}\alpha_iC_jM_{k\ell}\epsilon^{\ell mn}\mathbb{B}_{mn}\nonumber\\
& +C_i\mathbb{F}^{ij}X_j =0\,,\label{cond1}
\end{align}
where we have used the shorthand $C\cdot\alpha=C_i\alpha^i$. As above, it is helpful to separately consider the terms depending on $X_i$, $\alpha_{\mathrm{t}}$ and $e_{\mathrm{t}}^i$.

The computation showing that the terms proportional to $X_i$ in \eqref{cond1} cancel is straightforward. The only hint worth mentioning here is to make use of the identity
\[
\alpha^\ell\alpha_\ell\epsilon_{ijk}C^iA^jX^k=(\alpha\cdot C)\epsilon_{ijk}A^iX^j\alpha^k-(\alpha\cdot A)\epsilon_{ijk}X^i\alpha^j C^k+(\alpha\cdot X)\epsilon_{ijk}C^iA^j\alpha^k\,,
\]
which can be derived from $\alpha_{[\ell}\epsilon_{ijk]}=0$.

\medskip

The terms proportional to $\alpha_{\mathrm{t}}$ in \eqref{cond1} can be written as
\begin{align*}
& -\epsilon^{ijk}\eta_i\mathbb{B}_{j\ell}S^{\ell}_{\phantom{\ell}k}+\epsilon^{ijk}\eta_i\mathbb{B}_{\ell j}S^{\ell}_{\phantom{\ell}k}-\epsilon^{ijk}(S_i^{\phantom{i}\ell}\eta_{\ell})\mathbb{B}_{jk}\hspace*{5.7cm}
\end{align*}

\vspace*{-.7cm}

\begin{align*}
& -2\epsilon^{ijk}\alpha_i\eta_j(\mathbb{B}_{k\ell}A^\ell)+2\epsilon^{ijk}\eta_iA_j(\alpha^\ell\mathbb{B}_{\ell k})+2(\eta\cdot A)\epsilon^{ijk}\alpha_i\mathbb{B}_{jk}+4\epsilon^{ijk}\eta_i(\mathbb{B}_j^{\phantom{j}\ell}\alpha_{\ell})A_k\\
& +2\mathbb{B}\epsilon^{ijk}\eta_iA_j\alpha_k-2(A\cdot\alpha)\epsilon^{ijk}\eta_i\mathbb{B}_{jk}-4\epsilon^{ijk}(\mathbb{B}_{i\ell}\eta^\ell)A_j\alpha_k+2\epsilon^{ijk}(\eta^\ell\mathbb{B}_{\ell i})A_j\alpha_k\,,
\end{align*}
where we have introduced the shorthand $\eta_i:=\alpha_{\mathrm{t}}C_i$.

\medskip

In order to see that the $S$-dependent terms cancel out we make use of $\epsilon^{[ijk}S_i^{\phantom{i}\ell]}=0$. Checking that the terms linear in $\alpha_i$ in the previous expression cancel out is a little bit more involved, so we leave some of the details for Appendix \ref{appendix_details}.

\medskip

Finally, we discuss the computation of the terms proportional to $e_{\mathrm{t}}^i$ in \eqref{cond1}. This is, by far, the longest computation necessary to complete the Hamiltonian analysis of the Euclidean self-dual action. The first step is to introduce the $e_{\mathrm{t}}^i$-proportional terms of $M_{ij}$ and $N_{ij}$ into \eqref{cond1} and simplify the result by using, in particular, $S_{ij}\alpha^j=2A_i+\epsilon_{ijk}A^j\alpha^k$. After doing this some terms involving $S_{ij}$ still remain. This is somewhat disturbing because this could force us to write everything in terms of $S_{ij}$ (rather than $A_i$) which would make the computations much longer. Fortunately,  by replacing $\mathbb{B}_{ij}-\mathbb{B}_{ji}$ by $\epsilon_{ijk}\epsilon^{k\ell m}\mathbb{B}_{\ell m}$, whenever possible, it is possible to show that all the $S$-dependent terms can be grouped in the expression
\begin{equation}\label{Sterms}
(CS\alpha)\epsilon_{ijk}e_{\mathrm{t}}^i\mathbb{B}^{jk}+(e_{\mathrm{t}}S\alpha)\epsilon_{ijk}C^i\mathbb{B}^{jk}-(e_{\mathrm{t}}\cdot C)S^{i\ell}\alpha_\ell \epsilon_{ijk}\mathbb{B}^{jk}\,,
\end{equation}
where we have used the shorthand notation $CS\alpha:=C^iS_{ij}\alpha^j$, $e_{\mathrm{t}}S\alpha:=e_{\mathrm{t}}^iS_{ij}\alpha^j$ and $e_{\mathrm{t}}\cdot C:=e_{\mathrm{t}}^i C_i$. As we see, the combination $S_{ij}\alpha^j$ appears in all the terms of \eqref{Sterms} so we can write it in terms of $A_i$ as explained above. The result is
\begin{align*}
  & \,\,2(C\cdot A)\epsilon_{ijk}e_{\mathrm{t}}^i\mathbb{B}^{jk}+(C\cdot e_{\mathrm{t}})(A\mathbb{B}\alpha)+(\alpha\cdot e_{\mathrm{t}})(C\mathbb{B}A)+(e_{\mathrm{t}}\cdot A)(\alpha \mathbb{B}C)-(e_{\mathrm{t}}\cdot A)(C\mathbb{B}\alpha ) \\
  & -(C\cdot e_{\mathrm{t}})(\alpha\mathbb{B}A)-(\alpha\cdot e_{\mathrm{t}})(A\mathbb{B}C)+2(e_{\mathrm{t}}\cdot A)\epsilon_{ijk}C^i\mathbb{B}^{jk}+(\alpha\cdot C)(e_{\mathrm{t}}\mathbb{B}A)+(A\cdot C)(\alpha\mathbb{B}e_{\mathrm{t}}) \\
  & -(C\cdot A)(e_{\mathrm{t}}\mathbb{B}\alpha)-(\alpha\cdot C)(A\mathbb{B}e_{\mathrm{t}})-2(e_{\mathrm{t}}\cdot C)\epsilon_{ijk}A^i\mathbb{B}^{jk}\,,
\end{align*}
where, as in previous instances, we have used some self-explanatory notation. By adding this to the rest of the terms (the ones that can be directly written in terms of $A_i$) the final result for the terms proportional to $e_{\mathrm{t}}^i$ in \eqref{cond1} is
\begin{align*}
& \,\,3(e_{\mathrm{t}}\cdot A)\epsilon_{ijk}C^i\mathbb{B}^{jk}-2\epsilon_{ijk}C^ie_{\mathrm{t}}^j(\mathbb{B}^{k\ell}A_\ell)-3(e_{\mathrm{t}}\cdot C)\epsilon_{ijk}A^i\mathbb{B}^{jk}\hspace*{16mm}\\
&+2\mathbb{B}\epsilon_{ijk}e_{\mathrm{t}}^iA^jC^k+3\epsilon_{ijk}A^iC^j(e_{\mathrm{t}\ell}\mathbb{B}^{\ell k})+2\epsilon_{ijk}e_{\mathrm{t}}^i(\mathbb{B}^{j\ell}C_\ell)A^k-5\epsilon_{ijk}(\mathbb{B}^{i\ell}e_{\mathrm{t}\ell})A^jC^k\hspace*{-3mm}
\end{align*}

\vspace*{-.7cm}

\begin{align*}
&+2(e_{\mathrm{t}}\cdot A)\epsilon_{ijk}\alpha^iC^j(\mathbb{B}^{k\ell}\alpha_\ell)+2(\alpha \mathbb{B}\alpha)\epsilon_{ijk}e_{\mathrm{t}}^iC^jA^k-3(C\cdot \alpha)\epsilon_{ijk}A^i e_{\mathrm{t}}^j(\mathbb{B}^{k\ell}\alpha_\ell)\\
&+2(C\cdot e_{\mathrm{t}})\epsilon_{ijk}(\mathbb{B}^{i\ell}\alpha_\ell)A^j\alpha^k-(A\cdot C)(e_{\mathrm{t}}\cdot\alpha)\epsilon_{ijk}\alpha^i\mathbb{B}^{jk}-(C\mathbb{B}\alpha)\epsilon_{ijk}e_{\mathrm{t}}^iA^j\alpha^k\\
&+\epsilon_{ijk}e_{\mathrm{t}}^iA^j\alpha^k(\alpha\mathbb{B}C)+2(A\cdot\alpha)\epsilon_{ijk}C^ie_{\mathrm{t}}^j(\mathbb{B}^{k\ell}\alpha_\ell)+(A\cdot\alpha)(e_\mathrm{t}\cdot C)\epsilon_{ijk}\alpha^i\mathbb{B}^{jk}\\
&+(C\cdot\alpha)\epsilon_{ijk}e_{\mathrm{t}}^i(\alpha_{\ell}\mathbb{B}^{\ell j})A^k+\alpha^2\epsilon_{ijk}e_{\mathrm{t}}^i(\mathbb{B}^{j\ell}C_\ell)A^k-\alpha^2\epsilon_{ijk}e_{\mathrm{t}}^i(C_\ell \mathbb{B}^{\ell j})A^k\\
& +2(e_{\mathrm{t}}\mathbb{B}\alpha)\epsilon_{ijk}\alpha^iA^jC^k\,,
\end{align*}
where it is worth pointing out that all the linear, cubic and quartic terms in $\alpha_i$ are zero as a consequence of direct cancellations. This last expression can be shown to vanish. As in previous instances we leave some details about how this happens for Appendix \ref{appendix_details}.

%
%
\section{Hamiltonian formulation: Ahstekar variables and the time gauge}\label{sec_summary}

The Hamiltonian formulation obtained after completing the GNH procedure is formulated in a manifold $\mathsf{M}_0$ spanned by the fields $(e_{\mathrm{t}},e^i,\omega_{\mathrm{t}},\omega^i,\alpha_{\mathrm{t}},\alpha)$. The vector fields in this manifold have components $\mathbb{Y}_0=(Y_{e_{\mathrm{t}}}^i, Y_e^i, Y_{\omega_{\mathrm{t}}}^i, Y_\omega^i, Y_{\alpha_{\mathrm{t}}}, Y_\alpha)$.

By using the standard notation for 2-forms in field spaces the presymplectic 2-form on $\mathsf{M}_0$ can be written as
\begin{equation}\label{omega_canonical_transf}
\omega=\int_\Sigma \mathrm{d\!l}\omega^i{\wedge\!\!\!\wedge}\,\mathrm{d\!l}\left(\frac{1}{2}\epsilon_{ijk}e^j\wedge e^k+e_i\wedge \alpha\right)\,,
\end{equation}
which acting on vector fields $\mathbb{Y}\,,\mathbb{Z}$ in $\mathsf{M}_0$ gives \eqref{pullback_omega}.
The secondary constraints can be written as
\begin{align*}
&\epsilon_{ijk}e^j\wedge F^k+\alpha\wedge F_i=0\,,\\
&D\left(\frac{1}{2}\epsilon_{ijk}e^j\wedge e^k+e_i\wedge\alpha\right)=0\,,\\
&e_ i\wedge F^i=0\,,
\end{align*}
and the Hamiltonian vector fields are
\begin{align*}
  & Z_e^i=De_{\mathrm{t}}^i-\epsilon^{ijk}\omega_{\mathrm{t}j}e_k+E^i_{\phantom{i}j}e^j\,, \\
  & Z_\omega^i=D\omega_{\mathrm{t}}^i+W^i_{\phantom{i}j}e^j\,, \\
  & Z_\alpha\quad\, \mathsf{arbitrary}\,, \\
  & Z_{e_{\mathrm{t}}}^i\quad \mathsf{arbitrary}\,, \\
  & Z_{\omega_{\mathrm{t}}}^i\quad \mathsf{arbitrary}\,,  \\
  & Z_{\alpha_{\mathrm{t}}}\quad \mathsf{arbitrary}\,,
\end{align*}
with $W_{ij}$ given by \eqref{Wij},\eqref{Mij} and $E_{ij}$ by \eqref{Eij},\eqref{Nij}.

\medskip

The forms of the pullback of the symplectic form \eqref{omega_canonical_transf} and the fiber derivative \eqref{FL} strongly suggest the introduction of the object
\[
H_i:=\frac{1}{2}\epsilon_{ijk}e^j\wedge e^k+e_i\wedge \alpha\,,
\]
which would be canonically conjugate to $\omega_i$ in the sense that:
\[
\omega=\int_\Sigma \mathrm{d\!l}\omega^i{\wedge\!\!\!\wedge}\,\mathrm{d\!l}H_i\,.
\]
An important observation at this point is the following. Notice that the number of independent components in $H_i$ and $e_i$ are the same, hence it makes sense to write $e_i$ in terms of $H^i$ (or a suitably dualized object as we discuss below) to get a cleaner Hamiltonian description of Euclidean gravity. In fact, by proceeding in this way one arrives at the Ashtekar formulation for Euclidean gravity \textit{without having to use any gauge fixing}. This is in marked contrast with the situation in the case of the Holst action \cite{BarberoG:2020tit} and is significantly simpler.

To begin with it is convenient to define the following object
\begin{equation}\label{vect_density}
\vec{H}_i:=\frac{1}{\sqrt{1+\alpha^2}}\left(\frac{\cdot\wedge H_i}{\mathsf{vol}_e}\right)\,,
\end{equation}
that should be understood as an element of the double dual of the tangent space at each point of $\Sigma$. As the double dual of a finite dimensional vector space $V$ is canonically isomorphic to $V$, \eqref{vect_density} determines a unique vector field on $\Sigma$ that we also call $\vec{H}_i$. Given a 1-form $\beta\in\Omega^1(\Sigma)$ we have
\[
\imath_{\vec{H}_i}\beta=\frac{1}{\sqrt{1+\alpha^2}}\left(\frac{\beta\wedge H_i}{\mathsf{vol}_e}\right)\,.
\]
Taking this into account it is immediate to see that
\[
\imath_{\vec{H}_i}e^j=\frac{1}{\sqrt{1+\alpha^2}}(\delta_i^{\phantom{i}j}-\epsilon_{i}^{\phantom{k}jk}\alpha_k)\,.
\]
In the following it will be useful to introduce the 1-forms $h_i$ defined by
\[
h^i:=\frac{1}{\sqrt{1+\alpha^2}}\big(e^i+\alpha^i\alpha+\epsilon^{ijk}\alpha_je_k\big)\,.
\]
These satisfy the following important properties
\begin{align*}
& H_i=\frac{1}{2}\epsilon_{ijk}h^j\wedge h^k\,,\\
& \imath_{\vec{H}_i}h_j=\delta_{ij}\,,\\
& \mathsf{vol}_h=\sqrt{1+\alpha^2}\,\,\mathsf{vol}_e\,,
\end{align*}
where $\mathsf{vol}_h:=\frac{1}{3!}\epsilon_{ijk}h^i\wedge h^j\wedge h^k$ is a volume form on $\Sigma$. Notice that the last property implies that
\begin{equation}\label{vect_density_h}
\vec{H}_i:=\left(\frac{\cdot\wedge H_i}{\mathsf{vol}_h}\right)\,.
\end{equation}

We will now write the constraints in terms of $\omega_i$ and $\vec{H}_i$. First we compute
\begin{equation}\label{Asht_1}
  \imath_{\vec{H}_i}F^i=\frac{1}{\sqrt{1+\alpha^2}}(\mathbb{F}^{ij}\alpha_j-\epsilon^{ijk}\mathbb{F}_{jk}-\mathbb{F}^j_{\phantom{i}j}\alpha^i)e_i\,,
\end{equation}
and
\begin{equation}\label{Asht_2}
\epsilon^{ijk}\imath_{\vec{H}_i}\imath_{\vec{H}_j}F_k=\frac{2}{1+\alpha^2}\big(\alpha_i(\epsilon^{ijk}\mathbb{F}_{jk}-\mathbb{F}^{ij}\alpha_j)-\mathbb{F}^k_{\phantom{k}k}\big)\,.
\end{equation}
Obviously, the constraints \eqref{L1},\eqref{L3} in the form \eqref{L1_new},\eqref{L3_new} imply
\begin{align}
  & \imath_{\vec{H}_i}F^i=0\,,\label{L1_Ast}\\
  & \epsilon^{ijk}\imath_{\vec{H}_i}\imath_{\vec{H}_j}F_k=0\label{L3_Ast}\,,
\end{align}
which look very much like the vector and scalar constraints in the Ashtekar formulation for Euclidean gravity.

Conversely, the conditions \eqref{L1_Ast}-\eqref{L3_Ast} are equivalent to
\begin{align}
  & \mathbb{F}^{ij}\alpha_j-\epsilon^{ijk}\mathbb{F}_{jk}-\mathbb{F}^j_{\phantom{i}j}\alpha^i=0\,,\label{eq_1}\\
  & \alpha_i(\epsilon^{ijk}\mathbb{F}_{jk}-\mathbb{F}^{ij}\alpha_j)-\mathbb{F}^j_{\phantom{j}j}=0\label{eq_2}\,.
\end{align}
Multiplying the first of these two equations by $\alpha_i$ and adding it to the second leads to $(1+\alpha^2)\mathbb{F}^j_{\phantom{j}j}=0$ i.e. \eqref{L3_new}, and introducing this into \eqref{eq_1} we find \eqref{L1_new}.

Finally, in order to get the remaining constraint---equivalent to the usual Gauss law as will be shown later---we first compute (remember that for a vector field $X\in\mathfrak{X}(\Sigma)$ its divergence with respect to a given volume form $\mathsf{vol}$ is defined by $(\mathrm{div} X)\mathsf{vol}=\pounds_X\mathsf{vol}$ where $\pounds_X$ denotes the Lie derivative along $X$)
\[
\big(\mathrm{div}_h\vec{H}^i\big){\mathsf{vol}_h}:=\pounds_{\vec{H}^i}\mathsf{vol}_h=\mathrm{d}\imath_{\vec{H}^i}\mathsf{vol}_h=\mathrm{d}\left(\frac{1}{2}\epsilon_{ijk}h^j\wedge h^k\right)=\mathrm{d}H_i\,,
\]
whereas $ \epsilon_{ijk}\omega^j\wedge H^k$  can easily be seen to be equal to $\epsilon_{ijk}\big(\imath_{\vec{H}^k}\omega^j\big)\mathsf{vol}_h$. Hence, we conclude that the constraint \eqref{L2} written in the form $DH_i=0$ is equivalent to
\begin{equation}\label{Asht_3}
\mathrm{div}_h\vec{H}_i+\epsilon_{ijk}\imath_{\vec{H}^k}\omega^j=0\,.
\end{equation}
In order to make contact with the standard Ashtekar variables we introduce a fiducial volume form $\mathsf{vol}_0$ (which, if one wishes, may even be defined locally in terms of coordinates $x^i$, $i=1,2,3$) and write
\[
\omega=\int_\Sigma \mathrm{d\!l}\omega^i{\wedge\!\!\!\wedge}\,\mathrm{d\!l}H_i=\int_\Sigma \big(\mathrm{d\!l}\omega^i{\wedge\!\!\!\wedge}\,\mathrm{d\!l}\widetilde{H}_i\big)\mathsf{vol}_0\,.
\]
This should be understood as
\[
\omega(\mathbb{X},\mathbb{Y})=\int_\Sigma \big(Y_{\widetilde{H}_i}\iprod X_\omega^i-X_{\widetilde{H}_i}\iprod Y_\omega^i\big) \mathsf{vol}_0\,,
\]
with
\[
Y_{\widetilde{H}_i}:=\left(\frac{\cdot\wedge \epsilon_{ijk}h^j\wedge Y_h^k}{\mathsf{vol}_0}\right)
\]
and
\[
Y_{\widetilde{H}_i}\iprod\alpha=\left(\frac{\alpha\wedge \epsilon_{ijk}h^j\wedge Y_h^k}{\mathsf{vol}_0}\right)\,,
\]
for any 1-form $\alpha\in\Omega^1(\Sigma)$.

The relation between $\vec{H}_i$ and $\widetilde{H}_i$ is $\widetilde{H}_i=(\mathrm{det}\,h)\vec{H}_i$ with
\[
\mathrm{det}\,h:=\left(\frac{\mathsf{vol}_h}{\mathsf{vol}_0}\right)\,.
\]
In terms of $\widetilde{H}_i$ the constraints \eqref{L1_Ast} and \eqref{L3_Ast} can be immediately seen to be equivalent to
\begin{align}
  & \imath_{\widetilde{H}_i}F^i=0\,,\label{LL1_Ast}\\
  & \epsilon^{ijk}\imath_{\widetilde{H}_i}\imath_{\widetilde{H}_j}F_k=0\label{LL3_Ast}\,,
\end{align}
which are the vector and scalar constraints written in terms of the Ashtekar variables for Euclidean gravity.

As explained in  Appendix \ref{appendix_Gauss}, the constraint \eqref{Asht_3} in terms of $\widetilde{H}_i$ and $\omega_i$ becomes
\begin{equation}\label{LL2_Ast}
\mathrm{div}_0\widetilde{H}_i+\epsilon_{ijk}\imath_{\widetilde{H}^k}\omega^j=0\,,
\end{equation}
which, again, is exactly the usual Gauss law in the Ashtekar formulation. 

Another useful way to understand the constraints can be gained by introducing the objects
\[
^h\mathbb{F}_{ij}:=\left(\frac{F_i\wedge h_j}{\mathsf{vol}_h}\right)\,,\quad ^h\mathbb{B}_{ij}:=\left(\frac{Dh_i\wedge h_j}{\mathsf{vol}_h}\right)\,.
\]
In terms of them the constraints become
\begin{align*}
  & ^h\mathbb{F}_{[ij]}=0\,, \\
  & ^h\mathbb{B}_{[ij]}=0\,,\\
  & ^h\mathbb{F}_i^{\phantom{i}i}=0\,.
\end{align*}

Let us look now at the Hamiltonian vector fields in terms of the new variables. A direct computation using the definition of $h_i$ and $^h\mathbb{F}_{ij}$ gives the following expression for $Z_\omega^k$ on the final constraint submanifold (which means that we can make use of the constraints to simplify it)
\begin{equation}\label{eqZomegah}
  Z_\omega^k=D\omega_{\mathrm{t}}^k-\widehat{\alpha}_t\,\, ^h\mathbb{F}^k_{\phantom{k}\ell}h^\ell-\epsilon_{\ell mn}\widehat{e}_{\mathrm{t}}^{\,\,m} {\,\,^h\mathbb{F}}^{nk}h^\ell\,,
\end{equation}
with
\begin{equation}\label{redefparam}
\widehat{\alpha}_{\mathrm{t}}:=\frac{\alpha_{\mathrm{t}}-(e_{\mathrm{t}}\cdot\alpha)}{\sqrt{1+\alpha^2}}\,,\quad \widehat{e}_{\mathrm{t}}^{\,\,i}:=\frac{e_{\mathrm{t}}^i+\alpha_{\mathrm{t}}\alpha^i-\epsilon^{ijk}e_{\mathrm{t}j}\alpha_k}{\sqrt{1+\alpha^2}}\,.
\end{equation}
Although it is possible to get $Z_h^k$ by a direct, brute force approach, there is a much better---albeit slightly indirect---way to do it. The starting point is the identity
\[
\epsilon_{ijk}h^j\wedge Z_h^k=\epsilon_{ijk}e^j\wedge Z_e^k+Z_{ei}\wedge\alpha+e_i\wedge Z_{\alpha}\,,
\]
which comes directly from the definition of $H_i$. On the other hand, from \eqref{E2} we find
\begin{align*}
\epsilon_{ijk}e^j\wedge Z_e^k+Z_{ei}\wedge\alpha+e_i\wedge Z_{\alpha}&=\epsilon_{ijk}e^j\wedge De_{\mathrm{t}}^k+\omega_{\mathrm{t}}^j(e_j\wedge e_i)-\alpha\wedge De_{\mathrm{t}i}-\epsilon_{ijk}(\alpha\wedge e^j)\omega_{\mathrm{t}}^k\\
&+e_i\wedge \mathrm{d}\alpha_{\mathrm{t}}+e_{\mathrm{t}}^i\mathrm{d}\alpha+\epsilon_{ijk}e_{\mathrm{t}}^jDe^k-\alpha_{\mathrm{t}}De_i\,.
\end{align*}
By writing the terms on the right hand side of the previous expression in terms of $h_i$ and using \eqref{redefparam} we obtain the following equation for $Z_h^k$:
\[
\epsilon_{ijk}h^j\wedge Z_h^k=D\big(\epsilon_{ijk}\widehat{e}_{\mathrm{t}}^{\,\,j}h^k-\widehat{\alpha}_{\mathrm{t}}h_i\big)+\omega_{\mathrm{t}}^jh_j\wedge h_i\,.
\]
This equation involving differential forms can be solved by using the method explained in Appendix C of \cite{BarberoG:2021ekv}. The solution on the final constraint submanifold is
\begin{equation}\label{eqZh}
Z_h^k=D\widehat{e}_{\mathrm{t}}^{\,\,k}+\epsilon^k_{\phantom{k}\ell m}h^\ell\omega_{\mathrm{t}}^m-\frac{1}{2}\widehat{\alpha}_{\mathrm{t}}\,^h\mathbb{B}\, h^k-\epsilon_{\ell mn}\widehat{e}_{\mathrm{t}}^{\,\,m}\,^h\mathbb{B}^{nk}h^\ell+\epsilon^k_{\phantom{k}\ell m}\widehat{X}^m h^\ell +\widehat{\alpha}_{\mathrm{t}}\,\,^h\mathbb{B}^{k\ell}h_\ell\,,
\end{equation}
with
\[
\widehat{X}_i:=-\frac{1}{2}\epsilon_{ijk}\left(\frac{\mathrm{d}\widehat{\alpha}_{\mathrm{t}}\wedge h^j\wedge h^k}{\mathsf{vol}_h}\right)\,.
\]
At this point it is very interesting to compare the formulation that we have obtained in terms of the $h_i$ with the original one in the time gauge $\alpha=0$. The latter can be immediately obtained by substituting $\alpha=0$ in the pre-symplectic form \eqref{omega_canonical_transf}, the constraints \eqref{L1}-\eqref{L3} and the Hamiltonian vector fields $Z_\omega^k$, $Z_e^k$.

The role of a gauge fixing is to reduce (or eliminate) the arbitrariness due to the presence of arbitrary components in the Hamiltonian vector fields by fixing some or all of them. This can be done directly, or by demanding that the dynamics must be confined to a submanifold of $\mathsf{M}_0$ obtained by adding a gauge fixing condition to the secondary constraints. In the latter case we have to see what happens with the Hamiltonian vector fields. For the time gauge that we are considering here we must have $Z_\alpha=0$, or, equivalently, $X_\alpha=-\mathrm{d}\alpha_{\mathrm{t}}$.

The formulation in the time gauge can be summarized as follows: The presymplectic form is
\begin{equation}\label{omegaTG}
\omega=\int_\Sigma \mathrm{d\!l}\omega^i{\wedge\!\!\!\wedge}\,\mathrm{d\!l}\left(\frac{1}{2}\epsilon_{ijk}e^j\wedge e^k\right)\,,
\end{equation}
The secondary constraints become
\begin{align*}\label{consTG}
&\epsilon_{ijk}e^j\wedge F^k=0\,,\\
&D\left(\frac{1}{2}\epsilon_{ijk}e^j\wedge e^k\right)=0\,,\\
&e_ i\wedge F^i=0\,,
\end{align*}
or, equivalently,
\begin{align*}
  & \mathbb{F}_{[ij]}=0\,, \\
  & \mathbb{B}_{[ij]}=0\,,\\
  & \mathbb{F}_i^{\phantom{i}i}=0\,.
\end{align*}
Finally, the Hamiltonian vector fields are
\begin{align*}
Z_\omega^k&=D\omega_{\mathrm{t}}^k-{\alpha}_t\,\, \mathbb{F}^k_{\phantom{k}\ell}e^\ell-\epsilon_{\ell mn}{e}_{\mathrm{t}}^{\,\,m} {\,\,\mathbb{F}}^{nk}e^\ell\,,\\
Z_e^k&=D{e}_{\mathrm{t}}^{\,\,k}+\epsilon^k_{\phantom{k}\ell m}e^\ell\omega_{\mathrm{t}}^m-\frac{1}{2}{\alpha}_{\mathrm{t}}\,\mathbb{B}\, e^k-\epsilon_{\ell mn}{e}_{\mathrm{t}}^{\,\,m}\,\mathbb{B}^{nk}e^\ell+\epsilon^k_{\phantom{k}\ell m}{X}^m e^\ell +{\alpha}_{\mathrm{t}}\,\,\mathbb{B}^{k\ell}e_\ell\,,
\end{align*}
with
\[
{X}_i:=-\frac{1}{2}\epsilon_{ijk}\left(\frac{\mathrm{d}{\alpha}_{\mathrm{t}}\wedge e^j\wedge e^k}{\mathsf{vol}_e}\right)\,.
\]

\bigskip

As we can see a remarkable thing happens: the form of the presymplectic form, the constraints and the Hamiltonian vector fields obtained either by working with the $h_i$ variables or going to the time gauge in the original formulation \textit{is exactly the same} once we replace the arbitrary objects $\alpha_{\mathrm{t}}$ and $e_{\mathrm{t}}^i$ by the, also arbitrary, $\widehat{\alpha}_{\mathrm{t}}$ and $\widehat{e}_{\mathrm{t}}^{\,\,i}$. An interesting observation regarding this replacement of parameters is the fact that this comes from one of the $SO(3)$ factors of the $SO(4)$ symmetry of the action. Indeed, the infinitesimal transformations \eqref{SO3transformations_2} imply
\begin{align*}
\begin{split}
  & \delta_2\alpha_{\mathrm{t}}={\bm{\Upsilon}}_ie_{\mathrm{t}}^i\,,\\
  & \delta_2 e_{\mathrm{t}}^i=-{\bm{\Upsilon}}^i\alpha_{\mathrm{t}}+\epsilon^{i}_{\phantom{i}jk}e_{\mathrm{t}}^j{\bm{\Upsilon}}^k\,,
  \end{split}
\end{align*}
which is given by the matrix
\begin{equation*}
  \tau({\bm{\Upsilon}}_i)=
  \left[\begin{array}{rrrr}
  0& {\bm{\Upsilon}}_1& {\bm{\Upsilon}}_2&{\bm{\Upsilon}}_3\\
  -{\bm{\Upsilon}}_1 &0& {\bm{\Upsilon}}_3&-{\bm{\Upsilon}}_2\\
  -{\bm{\Upsilon}}_2 &-{\bm{\Upsilon}}_3&0& {\bm{\Upsilon}}_1\\
  -{\bm{\Upsilon}}_3 &{\bm{\Upsilon}}_2& -{\bm{\Upsilon}}_1&0\\
  \end{array}\right]\,.
\end{equation*}
The exponential of this matrix gives the matrix corresponding to a finite transformation
\begin{equation*}
 T({\bm{\Upsilon}})=\cos{\bm{\Upsilon}}\cdot \id_4+\frac{\sin {\bm{\Upsilon}}}{{\bm{\Upsilon}}}\cdot\tau({\bm{\Upsilon}})\,, \quad {\bm{\Upsilon}}:=\sqrt{{\bm{\Upsilon}}_1^2+{\bm{\Upsilon}}_2^2+{\bm{\Upsilon}}_3^2}\,.
\end{equation*}
If we write ${\bm{\Upsilon}}_i=-\frac{\alpha_i}{\alpha}\arctan\alpha$ with $\alpha:=\sqrt{\alpha_1^2+\alpha_2^2+\alpha_3^2}$, which is equivalent to $\alpha_i=-\frac{{\bm{\Upsilon}}_i}{{\bm{\Upsilon}}}\tan {\bm{\Upsilon}}$, with $\alpha_i\in(-1,1)$; the form of the previous finite transformation becomes
\begin{equation*}
  T(\alpha_i)=\frac{1}{\sqrt{1+\alpha^2}}
  \left[\begin{array}{cccc}
  1&-\alpha_1&-\alpha_2&-\alpha_3\\
  \alpha_1 &1& -\alpha_3&\alpha_2\\
  \alpha_2 &\alpha_3&1& -\alpha_1\\
  \alpha_3 &-\alpha_2&\alpha_1&1\\
  \end{array}\right]\,,
\end{equation*}
which gives \eqref{redefparam}.

\medskip

\noindent Several comments are in order now:

\begin{enumerate}
\item Both the 1-form $\omega_i$ and the 2-form $H_i$ are invariant under the transformations coming from \eqref{SO3transformations_2}. This immediately allows us to perform a \textit{symmetry reduction} and eliminate one of the two original $SO(3)$ symmetries form the final Hamiltonian formulation.

\item The primary constraint hypersurface $\mathsf{M}_0$ is spanned by  $(e_{\mathrm{t}},h^i,\omega_{\mathrm{t}},\omega^i,\alpha_{\mathrm{t}},\alpha)$ or equivalently by the set of fields obtained by replacing $h_i$ by $H_i$, $\vec{H}_i$ or $\widetilde{H}_i$. In the latter case we arrive at the usual Ashtekar formulation, but the other $h$-variables also provide interesting and equivalent phase space approaches to the dynamics of Euclidean GR.

\item  The field dynamics given by the vector fields obtained above, in particular that of $\omega_i$ and $h_i$ can be disentangled by introducing the uniquely defined vector field $\xi\in\mathfrak{X}(\Sigma)$ obtained by solving $\imath_\xi h^i=\widehat{e}_{\mathrm{t}}^{\,\,i}$ (uniqueness is a consequence of the non-degeneracy of the triads). A straightforward computation then gives
\[
Z_\omega^k=D(\omega_{\mathrm{t}}^k-\imath_\xi \omega^k)+\pounds_\xi\omega^k-\widehat{\alpha}_{\mathrm{t}}\,\,^h\mathbb{F}^{k\ell}h_\ell\,.
\]
As expected, a part of the dynamics corresponds to the infinitesimal diffeomorphisms defined by $\xi$ and $SO(3)$ gauge transformations parametrized by $\omega_{\mathrm{t}}^k-\imath_\xi \omega^k$. The non-trivial dynamics of Euclidean GR comes from the $-\widehat{\alpha}_{\mathrm{t}}\,\,^h\mathbb{F}^{k\ell}h_\ell$ term.

\item Proceding in a similar way one can rewrite $Z_h^k$ in a similar fashion and interpret part of the dynamics, again, as infinitesimal diffeomorphisms and local $SO(3)$ transformation given by the parameters $\xi$ and $\omega_{\mathrm{t}}^k-\imath_\xi \omega^k$. The cleanest way to see this is by introducing $Z_H^k$ as giving the dynamics of the 2-form $H_i$. In this case it is possible to show that
\[
Z_H^i=\epsilon^i_{\phantom{i}jk}H^j(\omega_{\mathrm{t}}^k-\imath_\xi \omega^k)+\pounds_\xi H^i-D(\widehat{\alpha}_{\mathrm{t}}h^i)\,,
\]
where $h^i$ should be written in terms of $H^i$. Notice that the GR dynamics is given here by the very simple term $-D(\widehat{\alpha}_{\mathrm{t}}h^i)$.

\item Finally, the dynamics of $\widetilde{H}^i$ can also be written in the same way. The result is exactly the one corresponding to the usual Ashtekar variables.

\item The rest of the fields $(e_{\mathrm{t}}$, $\omega_{\mathrm{t}}$, $\alpha_{\mathrm{t}}$, $\alpha)$ are arbitrary as there are no restrictions on the components of the Hamiltonian vector field giving their dynamics. This implies that we can choose them as any function of the dynamical fields and treat them as arbitrary external objects subject to the sole restriction of providing non-trivial dynamics for the system (for instance, $\widehat{\alpha}_{\mathrm{t}}$ should be different from zero everywhere on $\Sigma$).
\end{enumerate}

%
%
\section{Conclusions and comments}\label{sec_conclusions}

The main result of this paper is to show how the Ashtekar formulation for Euclidean gravity can be obtained from the self-dual action without introducing any gauge fixing (compare with reference \cite{BarberoG:2020tit}). This means that one can explicitly reduce the symmetry generated by one of the $SO(3)$ factors in $SO(4)=SO(3)\times SO(3)$ and remove the arbitrariness associated with the arbitrary components of the Hamiltonian vector fields that show up when using the GNH method. It should be clear at this point that, a completely analogous argument leading to the Ashtekar formulation without gauge fixing should apply in the case of using Dirac's approach. A secondary purpose of the paper is to complete the discussion of the consistency of the Hamiltonian formulation for Euclidean GR, which one of the authors has to admit, was not finalized in previous work on the subject starting from the action used here \cite{BarberoG:1994fcp}. In this sense it is instructive to compare the computations needed to complete the Hamiltonian analysis of the Holst action in the GNH framework \cite{BarberoG:2021ekv} with the ones presented here. Quite surprisingly for us, the complexity of the self-dual case is far greater than that of the case of using the Holst action. This is so even considering that the Hamiltonian analysis of the Holst action produces secondary constraints in two stages and not in one as it happens in the case analyzed in this paper.

In our opinion, the main use of the insights gained here may be to look for an action for Lorentzian GR which shares some of the nice features of the Euclidean self-dual action analyzed here. We hope that the clarification of the inner workings of the internal $SO(4)$ symmetry of the Euclidean model may help to better understand the much more relevant Lorentzian case.

%
%
\section*{Acknowledgments}

We would like to thank Wolfgang Wieland and Javier Olmedo for some interesting comments. This work has been supported by the Spanish Ministerio de Ciencia Innovaci\'on
y Universidades-Agencia Estatal de Investigaci\'on grant AEI/PID2020-116567GB-C22. E.J.S. Villase\~nor is supported by the Madrid Government (Comunidad
de Madrid-Spain) under the Multiannual Agreement with UC3M in the line
of Excellence of University Professors (EPUC3M23), and in the context of the
V PRICIT (Regional Programme of Research and Technological Innovation).

\begin{appendices}

%
%
\section{Tangency analysis: some details}\label{appendix_details}

\subsection{Additional details on the tangency condition \ref{tan3}}

The terms proportional to $e_{\mathrm{t}}^i$ and linear in $\alpha^i$ in this tangency condition are
\begin{align*}
  & \hspace*{-1mm}-4\epsilon_{ijk}e_{\mathrm{t}}^i(\mathbb{B}^{j\ell}\alpha_\ell)A^k-3(e_{\mathrm{t}}\cdot A)\epsilon_{ijk}\alpha^i\mathbb{B}^{jk}+3\epsilon_{ijk}\alpha^iA^j(e_{\mathrm{t}\ell}\mathbb{B}^{\ell k})+5\epsilon_{ijk}\alpha^i(\mathbb{B}^{j\ell}e_{\mathrm{t}\ell})A^k\label{alpha1}\\
  & \hspace*{-1mm}-2\mathbb{B}\epsilon_{ijk}\alpha^ie_{\mathrm{t}}^jA^k\!\!+\!2\epsilon_{ijk}\alpha^ie_{\mathrm{t}}^j(\mathbb{B}^{k\ell}A_\ell)\!+\!(e_{\mathrm{t}}\cdot\alpha)\epsilon_{ijk}A^i\mathbb{B}^{jk}
  \!\!+\!2(A\cdot\alpha)\epsilon_{ijk}e_{\mathrm{t}}^i\mathbb{B}^{jk}\!\!-\!2\epsilon_{ijk}(\alpha_\ell\mathbb{B}^{\ell i})e_{\mathrm{t}}^jA^k\,.\nonumber
\end{align*}
By using $\epsilon_{[ijk}\alpha_{\ell]}=0$ to transform the last term in the previous expression we get
\begin{align*}
  & -2\epsilon_{ijk}e_{\mathrm{t}}^i(\mathbb{B}^{j\ell}\alpha_\ell)A^k+3(e_{\mathrm{t}}\cdot\alpha)\epsilon_{ijk}A^i\mathbb{B}^{jk}-3(e_{\mathrm{t}}\cdot A)\epsilon_{ijk}\alpha^i\mathbb{B}^{jk}+3\epsilon_{ijk}\alpha^iA^j(e_{\mathrm{t}\ell}\mathbb{B}^{\ell k})\nonumber\\
  &-2\mathbb{B}\epsilon_{ijk}\alpha^ie_{\mathrm{t}}^jA^k+2\epsilon_{ijk}\alpha^ie_{\mathrm{t}}^j(\mathbb{B}^{k\ell}A_\ell)+5\epsilon_{ijk}\alpha^i(\mathbb{B}^{j\ell}e_{\mathrm{t}\ell}A^k)\,.
\end{align*}
Using now $\epsilon_{[ijk}e_{\mathrm{t}\ell]}=0$ to transform the last term in the previous expression we find
\begin{align*}
  & -2(e_{\mathrm{t}}\cdot\alpha)\epsilon_{ijk}A^i\mathbb{B}^{jk}-2\epsilon_{ijk}\alpha^iA^j(e_{\mathrm{t}\ell}\mathbb{B}^{\ell k})+2(e_{\mathrm{t}}\cdot A)\epsilon_{ijk}\alpha^i\mathbb{B}^{jk}-2\epsilon_{ijk}e_{\mathrm{t}}^i(\mathbb{B}^{j\ell}\alpha_\ell)A^k\\
  &+2\epsilon_{ijk}\alpha^ie_{\mathrm{t}}^j(\mathbb{B}^{k\ell}A_\ell)-2\mathbb{B}\epsilon_{ijk}\alpha^ie_{\mathrm{t}}^jA^k\,.
\end{align*}
By writing now the last term as $-2\mathbb{B}^\ell_{\phantom{\ell}\ell}\epsilon_{ijk}\alpha^ie_{\mathrm{t}}^jA^k$ and using $\mathbb{B}^\ell_{\phantom{\ell}[\ell}\epsilon_{ijk]}=0$ we get
\[
-2(e_{\mathrm{t}}\cdot\alpha)\epsilon_{ijk}A^i\mathbb{B}^{jk}+2(e_{\mathrm{t}}\cdot A)\epsilon_{ijk}\alpha^i\mathbb{B}^{jk}+2\epsilon_{ijk}A^i(\mathbb{B}^{j\ell}e_{\mathrm{t}\ell})\alpha^k-2\epsilon_{ijk}\alpha^iA^j(e_{\mathrm{t}\ell}\mathbb{B}^{\ell k})\,,
\]
which is zero as a consequence of $\epsilon_{[ijk}e_{\mathrm{t}\ell]}=0$.

\medskip

The terms  proportional to $e_{\mathrm{t}}^i$ and cubic in $\alpha^i$ are
\begin{align*}
  &+2(\alpha\mathbb{B}\alpha)\epsilon_{ijk}e_{\mathrm{t}}^iA^j\alpha^k+2\alpha^2\epsilon_{ijk}A^ie_{\mathrm{t}}^j(\mathbb{B}^{k\ell}\alpha_\ell)-2(e_{\mathrm{t}}\cdot\alpha)\epsilon_{ijk}(\mathbb{B}^{i\ell}\alpha_\ell)A^j\alpha^k\\
  &-2(A\cdot\alpha)\epsilon_{ijk}\alpha^ie_{\mathrm{t}}^j(\mathbb{B}^{k\ell}\alpha_\ell)\,.\hspace*{3.2cm}
\end{align*}
By writing now the first term in the previous expression as $2\mathbb{B}^{\ell m}\alpha_me_{\mathrm{t}}^iA^j\alpha^k \alpha_\ell\epsilon_{ijk}$ and using $\alpha_{[\ell}\epsilon_{ijk]}=0$ we immediately see that it cancels.

\subsection{Additional details on the tangency condition \ref{tan2}}

The terms proportional to $\alpha_{\mathrm{t}}$ (equivalently to $\eta_i$) in this tangency condition are
\begin{align*}
& -2\epsilon^{ijk}\alpha_i\eta_j(\mathbb{B}_{k\ell}A^\ell)+2\epsilon^{ijk}\eta_iA_j(\alpha^\ell\mathbb{B}_{\ell k})+2(\eta\cdot A)\epsilon^{ijk}\alpha_i\mathbb{B}_{jk}+4\epsilon^{ijk}\eta_i(\mathbb{B}_j^{\phantom{j}\ell}\alpha_{\ell})A_k\\
& +2\mathbb{B}\epsilon^{ijk}\eta_iA_j\alpha_k-2(A\cdot\alpha)\epsilon^{ijk}\eta_i\mathbb{B}_{jk}-4\epsilon^{ijk}(\mathbb{B}_{i\ell}\eta^\ell)A_j\alpha_k+2\epsilon^{ijk}(\eta^\ell\mathbb{B}_{\ell i})A_j\alpha_k\,,
\end{align*}
Using $\epsilon^{[ijk}\mathbb{B}_k^{\phantom{k}\ell]}=0$ we can transform the first term in the previous expression to get
\begin{align*}
& 2\epsilon^{ijk}\eta_iA_j(\alpha^\ell\mathbb{B}_{\ell k})+2(\eta\cdot A)\epsilon^{ijk}\alpha_i\mathbb{B}_{jk}+2\epsilon^{ijk}\eta_i(\mathbb{B}_j^{\phantom{j}\ell}\alpha_{\ell})A_k-2(A\cdot\alpha)\epsilon^{ijk}\eta_i\mathbb{B}_{jk}\\
& -2\epsilon^{ijk}(\mathbb{B}_{i\ell}\eta^\ell)A_j\alpha_k+2\epsilon^{ijk}(\eta^\ell\mathbb{B}_{\ell i})A_j\alpha_k\,.
\end{align*}
By using now $\epsilon^{[ijk}\eta^{\ell]}=0$ to transform the last term, the previous expression becomes
\begin{equation*}
2\epsilon^{ijk}\eta_iA_j(\alpha^\ell\mathbb{B}_{\ell k})+2\epsilon^{ijk}\eta_i(\mathbb{B}_j^{\phantom{j}\ell}\alpha_{\ell})A_k-2(A\cdot\alpha)\epsilon^{ijk}\eta_i\mathbb{B}_{jk}+2(\eta\cdot\alpha)\epsilon^{ijk}A_i\mathbb{B}_{jk}\,,
\end{equation*}
which vanishes as a consequence of $\epsilon^{[ijk}\alpha^{\ell]}=0$.

\medskip

The terms proportional to $e_{\mathrm{t}}^i$ and independent of $\alpha_i$ in the tangency condition are
\begin{align*}
& \,\,3(e_{\mathrm{t}}\cdot A)\epsilon_{ijk}C^i\mathbb{B}^{jk}-2\epsilon_{ijk}C^ie_{\mathrm{t}}^j(\mathbb{B}^{k\ell}A_\ell)-3(e_{\mathrm{t}}\cdot C)\epsilon_{ijk}A^i\mathbb{B}^{jk}\hspace*{16mm}\\
&+2\mathbb{B}\epsilon_{ijk}e_{\mathrm{t}}^iA^jC^k+3\epsilon_{ijk}A^iC^j(e_{\mathrm{t}\ell}\mathbb{B}^{\ell k})+2\epsilon_{ijk}e_{\mathrm{t}}^i(\mathbb{B}^{j\ell}C_\ell)A^k-5\epsilon_{ijk}(\mathbb{B}^{i\ell}e_{\mathrm{t}\ell})A^jC^k\,.
\end{align*}
Using $\epsilon_{[ijk}e_{\mathrm{t}\ell]}=0$ we can transform the last term in the previous expression to get
\begin{align*}
& -2\epsilon_{ijk}A^iC^j(e_{\mathrm{t}\ell}\mathbb{B}^{\ell k})-2(e_{\mathrm{t}}\cdot A)\epsilon_{ijk}C^i\mathbb{B}^{jk}+2(e_{\mathrm{t}}\cdot C)\epsilon_{ijk}A^i\mathbb{B}^{jk}+2\mathbb{B}\epsilon_{ijk}e_{\mathrm{t}}^iA^jC^k\\
& +2\epsilon_{ijk}e_{\mathrm{t}}^i(\mathbb{B}^{j\ell}C_\ell)A^k-2\epsilon_{ijk}C^ie_{\mathrm{t}}^j(\mathbb{B}^k_{\phantom{k}\ell}A^\ell)\,.
\end{align*}
By using now $\epsilon_{[ijk}\mathbb{B}^k_{\phantom{k}\ell]}=0$ to transform the last term we arrive at
\[
2\epsilon_{ijk}A^iC^j(\mathbb{B}^{k\ell}-\mathbb{B}^{\ell k})e_{\mathrm{t}\ell}-2(e_{\mathrm{t}}\cdot A)\epsilon_{ijk}C^i\mathbb{B}^{jk}+2(e_{\mathrm{t}}\cdot C)\epsilon_{ijk}A^i\mathbb{B}^{jk}\,,
\]
which can be seen to cancel because
\[
2\epsilon_{ijk}A^iC^j(\mathbb{B}^{k\ell}-\mathbb{B}^{\ell k})e_{\mathrm{t}\ell}=2\epsilon_{ijk}\epsilon^{k\ell m}\epsilon_{mpq}\mathbb{B}^{pq} A^iC^j=2(e_{\mathrm{t}}\cdot A)\epsilon_{ijk}C^i\mathbb{B}^{jk}-2(e_{\mathrm{t}}\cdot C)\epsilon_{ijk}A^i\mathbb{B}^{jk}\,.
\]

Finally, the terms proportional to $e_{\mathrm{t}}^i$ and quadratic in $\alpha_i$ in the tangency condition are
\begin{align*}
&+2(e_{\mathrm{t}}\cdot A)\epsilon_{ijk}\alpha^iC^j(\mathbb{B}^{k\ell}\alpha_\ell)+2(\alpha \mathbb{B}\alpha)\epsilon_{ijk}e_{\mathrm{t}}^iC^jA^k-3(C\cdot \alpha)\epsilon_{ijk}A^i e_{\mathrm{t}}^j(\mathbb{B}^{k\ell}\alpha_\ell)\\
&+2(C\cdot e_{\mathrm{t}})\epsilon_{ijk}(\mathbb{B}^{i\ell}\alpha_\ell)A^j\alpha^k-(A\cdot C)(e_{\mathrm{t}}\cdot\alpha)\epsilon_{ijk}\alpha^i\mathbb{B}^{jk}-(C\mathbb{B}\alpha)\epsilon_{ijk}e_{\mathrm{t}}^iA^j\alpha^k\\
&+\epsilon_{ijk}e_{\mathrm{t}}^iA^j\alpha^k(\alpha\mathbb{B}C)+2(A\cdot\alpha)\epsilon_{ijk}C^ie_{\mathrm{t}}^j(\mathbb{B}^{k\ell}\alpha_\ell)+(A\cdot\alpha)(e_\mathrm{t}\cdot C)\epsilon_{ijk}\alpha^i\mathbb{B}^{jk}\\
&+(C\cdot\alpha)\epsilon_{ijk}e_{\mathrm{t}}^i(\alpha_{\ell}\mathbb{B}^{\ell j})A^k+\alpha^2\epsilon_{ijk}e_{\mathrm{t}}^i(\mathbb{B}^{j\ell}C_\ell)A^k-\alpha^2\epsilon_{ijk}e_{\mathrm{t}}^i(C_\ell \mathbb{B}^{\ell j})A^k\\
& +2(e_{\mathrm{t}}\mathbb{B}\alpha)\epsilon_{ijk}\alpha^iA^jC^k\,.
\end{align*}
By using $e_{\mathrm{t}[\ell}\epsilon_{ijk]}=0$ we can transform the last term in the previous expression written in the form
\[
2(e_{\mathrm{t}}\mathbb{B}\alpha)\epsilon_{ijk}\alpha^iA^jC^k=2(\mathbb{B}^{\ell m}\alpha_m)e_{\mathrm{t}\ell}\epsilon_{ijk}\alpha^iA^jC^k
\]
to get
\begin{align*}
&\,\,2(\alpha \mathbb{B}\alpha)\epsilon_{ijk}e_{\mathrm{t}}^iC^jA^k-3(C\cdot \alpha)\epsilon_{ijk}A^i e_{\mathrm{t}}^j(\mathbb{B}^{k\ell}\alpha_\ell)-(A\cdot C)(e_{\mathrm{t}}\cdot\alpha)\epsilon_{ijk}\alpha^i\mathbb{B}^{jk}\\
&+2(e_{\mathrm{t}}\cdot\alpha)\epsilon_{ijk}A^iC^j(\mathbb{B}^{k\ell}\alpha_\ell)+\epsilon_{ijk}e_{\mathrm{t}}^iA^j\alpha^k(\alpha\mathbb{B}C)+2(A\cdot\alpha)\epsilon_{ijk}C^ie_{\mathrm{t}}^j(\mathbb{B}^{k\ell}\alpha_\ell)\\
&+(A\cdot\alpha)(e_\mathrm{t}\cdot C)\epsilon_{ijk}\alpha^i\mathbb{B}^{jk}+(C\cdot\alpha)\epsilon_{ijk}e_{\mathrm{t}}^i(\alpha_{\ell}\mathbb{B}^{\ell j})A^k+\alpha^2\epsilon_{ijk}e_{\mathrm{t}}^i(\mathbb{B}^{j\ell}C_\ell)A^k\\
&-\alpha^2\epsilon_{ijk}e_{\mathrm{t}}^i(C_\ell \mathbb{B}^{\ell j})A^k-(C\mathbb{B}\alpha)\epsilon_{ijk}e_{\mathrm{t}}^iA^j\alpha^k\,.
\end{align*}
The first term in this expression can be transformed by writing it in the form $2(\mathbb{B}^{\ell n}\alpha_n)\alpha_\ell\epsilon_{ijk}e_{\mathrm{t}}^iC^jA^k$ and using $\alpha_{[\ell}\epsilon_{ijk]}=0$. By doing this we get
\begin{align*}
& \,\,(C\cdot\alpha)\epsilon_{ijk}e_{\mathrm{t}}^i(\alpha_{\ell}\mathbb{B}^{\ell j})A^k+\epsilon_{ijk}e_{\mathrm{t}}^iA^j\alpha^k(\alpha\mathbb{B}C)-(C\cdot \alpha)\epsilon_{ijk}A^i e_{\mathrm{t}}^j(\mathbb{B}^{k\ell}\alpha_\ell)\\
&-(A\cdot C)(e_{\mathrm{t}}\cdot\alpha)\epsilon_{ijk}\alpha^i\mathbb{B}^{jk}+(A\cdot\alpha)(e_\mathrm{t}\cdot C)\epsilon_{ijk}\alpha^i\mathbb{B}^{jk}+\alpha^2\epsilon_{ijk}e_{\mathrm{t}}^i(\mathbb{B}^{j\ell}C_\ell)A^k\\
&-\alpha^2\epsilon_{ijk}e_{\mathrm{t}}^i(C_\ell \mathbb{B}^{\ell j})A^k-(C\mathbb{B}\alpha)\epsilon_{ijk}e_{\mathrm{t}}^iA^j\alpha^k\,.
\end{align*}
Now, we transform the first term by writing it as $\alpha^\ell e_{\mathrm{t}}^i(\alpha_n\mathbb{B}^{nj})A^kC_\ell\epsilon_{ijk}$ and using $C_{[\ell}\epsilon_{ijk]}=0$ to get
\begin{align*}
& (e_{\mathrm{t}}\cdot C)\epsilon_{ijk}(\alpha_\ell \mathbb{B}^{\ell i})A^j\alpha^k-(C\cdot \alpha)\epsilon_{ijk}A^i e_{\mathrm{t}}^j(\mathbb{B}^{k\ell}\alpha_\ell)-(A\cdot C)(e_{\mathrm{t}}\cdot\alpha)\epsilon_{ijk}\alpha^i\mathbb{B}^{jk}\\
&+(A\cdot C)\epsilon_{ijk}\alpha^ie_{\mathrm{t}}^j(\alpha_n \mathbb{B}^{nk})+(A\cdot\alpha)(e_\mathrm{t}\cdot C)\epsilon_{ijk}\alpha^i\mathbb{B}^{jk}-(C\mathbb{B}\alpha)\epsilon_{ijk}e_{\mathrm{t}}^iA^j\alpha^k\\
&+\alpha^2\epsilon_{ijk}e_{\mathrm{t}}^i(\mathbb{B}^{j\ell}C_\ell)A^k-\alpha^2\epsilon_{ijk}e_{\mathrm{t}}^i(C_\ell \mathbb{B}^{\ell j})A^k\,.
\end{align*}
In the next step we use $\epsilon_{[ijk}\alpha_{\ell]}=0$ to transform the first term and also rewrite the last two terms as
\[
\alpha^2\epsilon_{ijk}e_{\mathrm{t}}^i(\mathbb{B}^{j\ell}-\mathbb{B}^{\ell j})C_\ell A^k=\alpha^2(C\cdot A)\epsilon_{ijk}e_{\mathrm{t}}^i\mathbb{B}^{jk}-\alpha^2(e_{\mathrm{t}}\cdot C)\epsilon_{ijk}A^i\mathbb{B}^{jk}\,,
\]
this way we obtain
\begin{align*}
&-(C\cdot \alpha)\epsilon_{ijk}A^i e_{\mathrm{t}}^j(\mathbb{B}^{k\ell}\alpha_\ell)-(A\cdot C)(e_{\mathrm{t}}\cdot\alpha)\epsilon_{ijk}\alpha^i\mathbb{B}^{jk}+(e_{\mathrm{t}}\cdot C)\epsilon_{ijk}A^i\alpha^j(\mathbb{B}^{k\ell}\alpha_\ell)\\
&+(A\cdot C)\epsilon_{ijk}\alpha^ie_{\mathrm{t}}^j(\alpha_n \mathbb{B}^{nk})-(C\mathbb{B}\alpha)\epsilon_{ijk}e_{\mathrm{t}}^iA^j\alpha^k+\alpha^2(C\cdot A)\epsilon_{ijk}e_{\mathrm{t}}^i\mathbb{B}^{jk}\,.
\end{align*}
We can now transform the last term by writing it as $(C\cdot A)e_{\mathrm{t}}^i\mathbb{B}^{jk}\alpha^\ell \alpha_{\ell}\epsilon_{ijk}$ and using $\alpha_{[\ell}\epsilon_{ijk]}=0$. This way the previous expression becomes
\begin{align*}
&-(C\cdot \alpha)\epsilon_{ijk}A^i e_{\mathrm{t}}^j(\mathbb{B}^{k\ell}\alpha_\ell)+(e_{\mathrm{t}}\cdot C)\epsilon_{ijk}A^i\alpha^j(\mathbb{B}^{k\ell}\alpha_\ell)+(A\cdot C)\epsilon_{ijk}\alpha^ie_{\mathrm{t}}^j(\mathbb{B}^{k\ell}\alpha_\ell)\\
&-(C\mathbb{B}\alpha)\epsilon_{ijk}e_{\mathrm{t}}^iA^j\alpha^k=4e_{\mathrm{t}}^mA^i\alpha^j(\mathbb{B}^{k\ell}\alpha_\ell)\epsilon_{[ijk}C_{\ell]}=0\,.
\end{align*}

%
%
\section{The Gauss law in the Ashtekar formulation: some computational details}\label{appendix_Gauss}

By multiplying the constraint \eqref{Asht_3} by $\mathrm{det}\,h$ (which is non-zero at every point of $\Sigma$) we get the following equivalent expression:
\begin{equation}\label{apB_1}
  (\mathrm{det}\,h)\cdot\mathrm{div}_h\left(\frac{\widetilde{H}^i}{\mathrm{det}\,h}\right)+\epsilon_{ijk}\imath_{\widetilde{H}^k}\omega^j=0\,.
\end{equation}
We now prove that
\begin{equation}\label{apB_2}
(\mathrm{det}\,h)\cdot\mathrm{div}_h\left(\frac{\widetilde{H}^i}{\mathrm{det}\,h}\right)=\mathrm{div}_0\widetilde{H}^i\,,
\end{equation}
for any field independent volume form $\mathsf{vol}_0$ [remember that, for a given volume form $\mathsf{vol}$, the divergence of a vector field $X$ is defined as $\left(\frac{\pounds_X \mathsf{vol}}{\mathsf{vol}}\right)$]. 

In order to prove \eqref{apB_2} we need the identity
\begin{equation}\label{apB_3}
(\mathsf{vol}_0)\cdot\pounds_{\widetilde{H}^i}\mathrm{det}\,h=\mathrm{d}(\mathrm{det}\,h)\wedge \imath_{\widetilde{H}^i}\mathsf{vol}_0\,,
\end{equation}
which can be obtained by taking the interior product $\imath_{\widetilde{H}^i}$ of both sides of the trivial identity $0=\mathrm{d}(\mathrm{det}\,h)\wedge {\mathsf{vol}}_0$.

For $X\in\mathfrak{X}(\Sigma)$, $\varphi\in\Omega^0(\Sigma)$ and any volume form $\mathsf{vol}\in\Omega^3(\Sigma)$ we have
\[
\mathrm{div}(\varphi X)=\pounds_X\varphi+\varphi\, \mathrm{div}X\,.
\]
Using this expression one gets
\begin{equation}\label{apB_4}
\hspace*{-2mm}(\mathrm{det}\,h)\mathrm{div}_h\left(\frac{\widetilde{H}^i}{\mathrm{det}\,h}\right)=(\mathrm{det}\,h)\pounds_{\widetilde{H}^i}\left(\frac{1}{\mathrm{det}\,h}\right)\!+\mathrm{div}_h\widetilde{H}^i\!=
-\frac{1}{\mathrm{det}\,h}\pounds_{\widetilde{H}^i}\mathrm{det}\,h+\mathrm{div}_h\widetilde{H}^i\,.
\end{equation}
Finally, making use of \eqref{apB_3} we find
\begin{align*}
  \mathrm{div}_h\widetilde{H}^i&=\left(\frac{\pounds_{\widetilde{H}^i}{\mathsf{vol}}_h}{\mathsf{vol}_h}\right)=\left(\frac{\mathrm{d}[(\mathrm{det}\,h)\cdot (\imath_{\widetilde{H}^i}\mathsf{vol}_0)]}{\mathsf{vol}_h}\right)=\left(\frac{\mathrm{d}(\mathrm{det\,h})\wedge\imath_{\widetilde{H}^i}\mathsf{vol}_0}{\mathsf{vol}_h}\right)  \\
  &+\left(\frac{(\mathrm{det}\,h)\pounds_{\widetilde{H}^i}\mathsf{vol}_0}{\mathsf{vol}_h}\right)=\left(\frac{(\mathsf{vol}_0)\cdot\pounds_{\widetilde{H}^i}\mathrm{det}\,h}{\mathsf{vol}_h}\right)
  +\left(\frac{(\mathrm{det}\,h)\pounds_{\widetilde{H}^i}\mathsf{vol}_0}{\mathsf{vol}_h}\right)\\
  &=\frac{1}{\mathrm{det}\,h}\pounds_{\widetilde{H}^i}\mathrm{det}\,h+\left(\frac{\pounds_{\widetilde{H}^i}\mathsf{vol}_0}{\mathsf{vol}_0}\right)
  =\frac{1}{\mathrm{det}\,h}\pounds_{\widetilde{H}^i}\mathrm{det}\,h+\mathrm{div}_0\widetilde{H}^i\,,
\end{align*}
which introduced in \eqref{apB_4} gives \eqref{apB_2}\,.

\end{appendices}

%
%

\end{document}